\documentclass[aps,amsmath,amssymb,showpacs,showkeys]{revtex4} 
\usepackage[dvips]{graphicx,color}%
\usepackage{times}
\usepackage{multirow}
\usepackage{xcolor}
\usepackage[
  colorlinks=true,
  urlcolor=magenta,
  linkcolor=red,
  citecolor=blue]{hyperref}

\begin{document}
\title{Radial oscillations and gravitational wave echoes of strange stars for various equations of state}

\author{Jyatsnasree Bora}
\email[Email: ]{jyatnasree.borah@gmail.com} 

\author{Umananda Dev Goswami}
\email[Email: ]{umananda2@gmail.com}

\affiliation{Department of Physics, Dibrugarh University, Dibrugarh 786004, 
Assam, India}

\begin{abstract}
We study the radial oscillations of non-rotating strange stars and their 
characteristic echo frequencies for three Equations of state (EoSs), viz., MIT 
Bag model EoS, linear EoS and polytropic EoS. The frequencies of radial 
oscillations of these compact stars are computed for these EoSs. 22 lowest 
radial frequencies for each of these three EoSs have been computed. First, for 
each EoS, we have integrated Tolman-Oppenheimer-Volkoff (TOV) equations 
numerically to calculate the radial and pressure perturbations of strange 
stars. Next, the mass-radius relationships for these stars are obtained using 
these three EoSs. Then the radial frequencies of oscillations for these EoSs 
are calculated. Further, the characteristic gravitational wave echo frequencies
and the repetition of echo frequencies of strange stars are computed for these 
EoSs. Our numerical results show that the radial frequencies and also echo 
frequencies vastly depend on the model and on the value of the model 
parameter. Our results also show that, the radial frequencies of strange stars  
are maximum for polytropic EoS in comparison to MIT Bag model EoS and linear 
EoS. Moreover, strange stars with MIT Bag model EoS and linear EoS are found to 
emit gravitational wave echoes. Whereas, strange stars with polytropic EoS are 
not emitting gravitational wave echoes.   
\end{abstract}


\maketitle


\section{Introduction}
One of the most interesting areas of study in astrophysics and cosmology is 
the area of strange matter objects \citep{Farhi1984, Witten1984, Aclock1988}. Various 
authors have been working in this area for a long time, mostly trying to 
constrain mass and radius profiles of stars formed from strange matter using 
different Equations of state (EoSs) \citep{overgard1991, zdunik2002, mart2010, 
Weber2012}, which are yet to be known exactly. The astrophysical compact objects, 
which are entirely made of quark matter or strange matter (deconfined $u$, $d$, $s$ 
quark matter), are known as Strange Stars (SSs) \citep{Aclock1988, weber2005}. These 
compact stars with extremely high densities are unique probes to study the properties 
of matter under extreme conditions. The study of such stars (i.e.\ SSs) has 
appeared as a subject of considerable interest over the last few decades due to 
their unusual internal composition, structure and physical properties as 
compared to other compact objects, like White Dwarf (WD) and Neutron Star (NS) 
\citep{shapiro2004}. An indirect approach to study such highly compact stars is 
asteroseismology \citep{handler2012}. By observing the radial and non-radial modes 
of oscillations of a star, one can reveal the stability condition, mass, 
radius, composition, etc.\ of the star. Though the radial mode, where a regular 
change in shape and size of the oscillating body occurs, is the simplest 
oscillation mode, yet it is a great tool for collecting information about the 
star. The very first study on such radial mode of oscillations of compact stars 
were carried out by 
\citep{Chandrasekhar1964a, Chandrasekhar1964b}. After that, radial modes have been 
investigated for compact stars, like WD and NS for various nuclear matter EoSs 
by others \citep{Chanmugam1977, Glass1983, haensel1989}. In the paper 
\citep{Chanmugam1977}, the radial oscillations of zero-temperature degenerate stars 
(WD and NS) were calculated for the fundamental and first two excited modes. Using 
the polytropic model EoS, P.\ Haensel \textit{et al.}\ studied the pulsation 
properties of NS undergoing a phase transition to quark matter \citep{haensel1989}. 
For SSs, using the general relativistic pulsation equation, the 
eigenfrequencies of radial pulsations were calculated by B.\ Datta 
\textit{et al.}\ in 1992 \citep{datta1992}. They also calculated the oscillation time 
periods. The calculation of radial oscillations of SS and NS for 2 
lowest order oscillation modes were reported in \cite{vath1992}. In recent 
studies, G.\ Panotopoulos and I.\ Lopes computed radial oscillation modes for 
SS admixed with dark matter \citep{panotopoulos2017,panotopoulos2018}. Study of the 
anisotropic SS in non-linear EoS was reported by I.\ Lopes in \citep{lopes2019}. The 
displacements in such radial oscillations can be described mathematically by Sturm-
Liouville type differential equations, which yield discrete eigensolutions: the 
radial mode frequencies of the given model \citep{handler2012}. A new numerical 
algorithm to solve Sturm-Liouville differential equations governing the radial 
oscillations of non-rotating SS (see Sec.\ \ref{radial}) is reported 
in \citep{clemente2020}. 

An interesting property of such ultracompact objects is that some of them can 
reflect or echo the Gravitational Waves (GWs) signal \citep{ferrari2000, pani2018}. 
In 1978, Yu.\ G.\ Ignat'ev and A.\ V.\ Zakharov showed that GW can be reflected 
from sufficiently dense stellar bodies \citep{ignatev1978}. They calculated the 
condition of reflection and reflection index for NSs. Recently more authors 
\citep{abedi2017, abedi2019, abedi2020, conklin2018} have reported the possibility 
of GW Echo (GWE) from the ultracompact objects formed in the merging of massive 
objects, like Black Holes (BHs) and NSs. The echo signal originating from 
ultracompact star was first reported in \citep{pani2018}. For non-spinning, 
constant-density stars they have calculated the characteristic echo frequency $
\approx 72$ Hz at $4.2\sigma$ significance level. Those ultracompact massive 
post-merger objects formed in GW generation events, which feature a photon sphere can 
only echo the GWs or can trap the GWs partially. The photon sphere is a surface over 
a massive star located at $R=3M$, where $R$ is the radius and $M$ is the mass of 
the star \citep{mannarelli2018}. In photon sphere, circular photon orbits are 
possible and are featured by both BHs and ultracompact stars 
\citep{mannarelli2018}. Thus the emission of GWE takes place if and only if a star 
features a photon sphere and also it should be very compact, close to the 
Buchdhal's limit radius $R_{B}=9M/4$ \citep{buchdahl}. Thus for GWE, the 
compactness of the final compact object should lie in between $1/3$ (to have a 
photon sphere) and $4/9$ (to emit GWEs at a frequency of tens of Hertz) 
\citep{pani2018, mannarelli2018}. In the recent GW merging event GW170817, the nature 
of final supermassive stellar object formed is not clearly established 
\citep{abbott}. Considering this final supercompact object as a SS, M. Mannareli 
and F. Tonneli calculated the corresponding echo frequency using the MIT Bag 
model EoS \citep{mannarelli2018}. The authors have explained the reason for 
considering the final compact remnant as a SS in their paper. For the 
considered model of SS with $B_{1}=(145\, \mbox{MeV})^{4}$ and $B_{2}=(185\,
\mbox{MeV})^{4}$, they have calculated the echo frequency as $\omega_{1,echo}
=17$ kHz and $\omega_{2,echo}=27$ kHz.

Based on some possible evidences on the existence of SSs \citep{li, henderson}, 
and also motivated by previous works mentioned above, in this work we have 
examined the radial oscillation frequencies and GWE frequencies of SSs for 
three EoSs, viz., MIT Bag model EoS, linear EoS and polytropic EoS. As the 
external properties of SSs crucially depend on EoSs, so the oscillation and 
echo frequencies will be different for different EoSs. Thus the use of three 
EoSs in this study will enable us to compare the results obtained from three 
EoSs, which may be useful to constrain these models in the future from the 
observational data of SSs.   

The rest of the paper is organized as follows: In Sec.\ \ref{eos} we have 
introduced the EoSs that are considered in this work. In Sec.\ \ref{radial}, 
the equations governing the radial oscillations, i.e.\ the hydrostatic 
equilibrium Tolman-Oppenheimer-Volkoff (TOV) equations are briefly summarized, 
and the equations for radial and pressure perturbations are 
discussed. In Sec.\ \ref{echo}, a brief discussion on GWE frequencies is 
made. The numerical results are given in Sec.\ \ref{numerical} and finally we 
conclude the paper in Sec.\ \ref{conclusion}. In this work we consider the 
natural unit system, in which $ c = \hbar = 1$, i.e. all dimensionful 
quantities are measured in GeV. Also we assume $G=1$ and adopt the metric 
convention: $(-,\,+,\,+,\,+)$.
 

\section{Equation of States}\label{eos}
The physics of very high density matter, like SS matter is still not pretty 
clear till date. In order to construct a compact star's model, an EoS has to 
be specified, which is the relation between the pressure $p$ and the energy 
density $\rho$. A definite EoS for a compact object can give properties, like 
mass, the mass-radius relationship, the crust thickness, the cooling rate, etc. 
In most of the studies on SSs, the most simple MIT Bag model EoS framework is 
used. As mentioned in the previous section, we have considered three EoSs, 
viz., the MIT Bag model, linear and polytropic EoSs in order to describe SSs. 
These three EoSs can be used to describe the structure of SSs in General 
Relativity (GR). The MIT Bag model EoS is the simplest EoS corresponding to a 
relativistic gas of deconfined quarks with energy density \citep{haensel}, and in 
this model it is considered that a universal pressure, known as the Bag 
constant, on the surface of any region containing quarks causes the quark 
confinement \citep{chodosa, sharma}. According to this model, the stiffer 
EoS of the strange matter has a simple linear form as given by 
\citep{mannarelli2018} 
\begin{equation}
\label{eq1}
p = \rho - 4\,B, 
\end{equation}
where $\rho$ is the energy density, $p$ is the isotropic pressure and $B$ is 
the Bag constant. The original form of this model of hadron structure was
presented in 1974 by \citep{chodosa, chodosb} and used by \citep{Witten1984} 
in his calculation of SS mass-radius relation \citep{alcock}. For SSs in 
the MIT Bag model EoSs, the maximum mass $M_{max}$ and radius $R$ vary with the 
Bag constant $B$ as $\propto B^{-1/2}$ \citep{Witten1984}. In our study we 
have considered three values of $B$ as $(190\, \mbox{MeV})^{4}$, 
$(217\, \mbox{MeV})^{4}$ and $(243\, \mbox{MeV})^{4}$. These values of 
$B$ are chosen here in order to get physically motivated SS configurations 
in the sense that they almost lie in the acceptable range of values of $B$ 
as suggested by \citep{aziz, carinhas}. However, it should be noted that 
till now no stringent physically acceptable range of $B$ is available. 
Moreover, this model EoSs give SS configrations with a constant compactness 
independent of $B$, and within the range of $B$ of our interest the required 
criterion of maximum mass, radius and compactness for the emission of GWEs 
are fulfilled.

Besides the simple MIT Bag model EoS, Dey \textit{et al.}\ in 1998 \citep{dey} 
developed another model for SSs, in which an interquark vector potential and a 
density based scalar potential describe the quark interactions \citep{sharma}. 
The interquark vector potential is originated from gluon exchange and the 
density dependent scalar potential is used to restore chiral symmetry at a high 
density. The EoS obtained from this model can be approximated to a linear form, 
known as the linear EoS, and is given as
\begin{equation}
\label{eq2}
p = b \,(\rho - \rho_{s}),
\end{equation}
where $b$ and $\rho_{s}$ are two model parameters, specifically $\rho_s$ 
represents the surface energy density and $b$ is a linear constant 
\citep{rosinska}. In particular, we have used three values of linear constant, 
which are $b=0.910$, $0.918$ and $0.926$, and the corresponding values of 
surface energy density $\rho_{s}$ are taken in this study. For this EoS, 
the maximum mass and corresponding radius of SS increases slowly with 
increasing $b$ value. While choosing the said values of $b$ we have kept in 
mind the conditions for emitting GWE frequency from SS, which impose the 
restriction that compactness of SS should lie within $0.33$ and $0.44$. Under
this restriction and the condition for causality $0.710$ is the minimum and $1.000$ is the maximum
values of $b$, and so our chosen values of $b$ are well within the allowed 
compactness range of SSs emitting GWEs.

It is also possible to describe compact stars using polytropic EoS. Hence,
in this work we have used also the following polytropic EoS to investigate the 
behaviour of SSs in GR \citep{tooper1, tooper2, thiru, Herrera, kokkotas}: 
\begin{equation}
\label{eq3}
p = k\,\rho^{\,\Gamma}, 
\end{equation}
where $k$ is the polytropic constant, $\Gamma$ is the polytropic exponent with 
$\Gamma=1+1/n$, $n$ being the polytropic index. This EoS is one of the 
primitive polytropic relation describing compact stars. It should be noted 
that a variation in polytropic index is closely related to different stellar 
structures. Here we have chosen three different values of polytropic 
exponent, viz., $1.5$, $1.67$ and $2$ as guess values to describe the 
structure of SSs. The polytropic constant $k$ depends on central pressure and 
density of compact stars \citep{tooper2}. Thus the choice of central density 
and central pressure highly influences the value of k. For $\Gamma=1.5$, the 
exact solutions for relativistic polytropes with a polytropic EoS have been 
found by Thirukkanesh and Ragel \citep{thiru}. Their study shows that a 
polytropic 
compact object with $n=2$ ($\Gamma=1.5$) is viable to experimental results. 
For $\Gamma=1.5$ we have chosen $k=7.6\times10^{38}$ corresponding to a 
suitable central density and central pressure of SSs. This value of $k$ is
obtained by setting $c=\hbar=G=1$ and using the relation: $1$ GeV = 
$1 M_{P}/(1.22\times10^{19})$, where $M_{P}$ is the Planck mass, which is taken
as unity in this work. Values of $k$ for other two $\Gamma$s are calculated
accordingly. Thus, for $\Gamma=1.67$ corresponding to $n=1.5$, $k=0.05 \, 
{\mbox{fm}^{8/3}} \approx 1.7\times10^{51}$ is taken \citep{ray}. We have 
chosen some higher value of $\Gamma$, i.e.\ $\Gamma=2$ corresponding to $n=1$ 
and $k = 1.1 \times 10^{-4} \, {\mbox{fm}^{3}} \, {\mbox{MeV}^{-1}} 
\approx 4.6\times10^{77}$ \citep{lattimer}. As it is well known that compact 
stars, like neutron stars are best described for the polytropic exponent 
$\Gamma$s lying in the range of $2$ and $3$ \citep{lattimer, kokkotas, ns}. So, 
in order to describe SSs we have chosen the lower value of polytropic 
exponent, i.e.\ $\Gamma=2$ from this range. The value $\Gamma = 1.67$ is taken
as intermediate value between $1.5$ and $2$ for the corresponding values of 
$n$.   

Once the EoS of a star is known, the TOV equations can be integrated 
numerically to calculate the macroscopic features of the star, such as 
its mass and radius.


\section{Radial oscillations of strange stars}\label{radial}
To study a relativistic star in 
GR one needs to solve first the Einstein's field equations for a given 
spacetime metric. The Einstein's field equations in a compact form is given by 
\begin{equation}
	\label{eq4}
	G_{\mu \nu} = R_{\mu \nu} - \dfrac{1}{2} g_{\mu \nu} R = 8 \pi T_{\mu 
	\nu},
	\end{equation}  
where $G_{\mu \nu}$ is the Einstein tensor, $g_{\mu \nu}$ is the metric tensor,
$T_{\mu \nu}$ is the energy-momentum tensor, $R_{\mu \nu}$ is the Ricci
tensor and $R$ is the Ricci scalar. For a spherically symmetric static system 
or stellar object, we have to use the Schwarzschild metric, given as 
	\begin{equation}
	\label{eq5}
	ds^{2} = -\,e^{\chi(r)}dt^{2} + e^{\lambda(r)}dr^{2} + r^{2}\,d\Omega^{2},
    \end{equation}
where as usual $t$ and $r$ respectively represent the time and space 
coordinates and $d\Omega^{2}=d\theta^{2}+\sin^{2}\theta\, d\phi^{2}$; 
$\theta$ and $\phi$ are polar and azimuthal angles. The metric parameters 
$\chi$ and $\lambda$ are functions of $r$ only and can be replaced by
	\begin{equation}\label{eq6}
	e^{-\,\chi} = e^{\lambda} = \dfrac{1}{1-2M/r},
	\end{equation}
where $M$ is the mass at the radius $R$ of the star and $r<R$. 
In GR, when a star is in hydrostatic equilibrium then its interior structure 
is described by the TOV equations \citep{tolman,tov}, which are derived by 
solving the Einstein's Eqs.\ (\ref{eq4}) for the Schwarzschild metric 
(\ref{eq5}). So, for simplicity if we neglect the rotation of the compact 
stellar objects, then the structure of such objects can be obtained by solving 
the TOV equations, which are
	\begin{align}
	\label{eq7}
	\dfrac{d\chi}{dr} =& \;- \dfrac{2}{\rho + p} \dfrac{dp}{dr},\\[5pt]
	\label{eq8}
	\dfrac{dm}{dr} =&\; 4\pi \rho\, r^{2},\\[5pt]
	\label{eq9}
	\dfrac{dp}{dr} =&\;- (\rho + p)\left(\frac{m}{r^2}+4\pi p\,r\right)\left(1 
	- \frac{2m}{r}\right)^{-1},
	\end{align}
where $\rho=\epsilon/c^{2}$ is the energy density, c being the speed of light 
in vacuum which is considered to be equal to unity in this work.

The set of hydrostatic equilibrium equations is an initial value problem and 
can be solved numerically. Now for some given EoSs, the TOV Eqs.\ (\ref{eq8}) 
and (\ref{eq9}) are to be integrated with initial conditions : (i) mass at the 
centre is zero, i.e. $m(r=0)=0$ and (ii) pressure at the centre 
becomes the central pressure, i.e. $p(r=0)=p_{c}$. Moreover, the radius of the 
star can be determined by the fact that the energy density vanishes at the 
surface, i.e. $\rho (r=R)=0$. At the surface of the star the mass can be given 
as $m(r=R)=M$. And finally the metric term $\chi$ can be calculated from 
Eq.\ (\ref{eq7}) using the boundary condition:
\begin{equation}\label{eq10}
\chi(R)=\ln\,(1-\dfrac{2M}{R}).
\end{equation}
The solution of TOV equations for a particular star will give the information 
about the star and using these information the asteroseismic behaviour of that
star can be revealed. 

In 1964, S.\ Chandrasekhar for the first time established equations that govern
the radial oscillations of a gas sphere in the general relativistic framework 
taking into account radial and pressure perturbations 
\citep{Chandrasekhar1964a,Chandrasekhar1964b}. Defining the dimensionless variables $\xi= 
\Delta r/r$ and $\eta = \Delta p/p$, where $\Delta r$ is the radial 
perturbation and $\Delta p$ is the corresponding Lagrangian perturbations of 
the pressure, the Chandrasekhar's equations of radial and pressure 
perturbations can be written as \citep{Chandrasekhar1964a, Chandrasekhar1964b, Chanmugam1977, vath1992, panotopoulos2017}
\begin{equation}
	\label{eq11}
	\small {\dfrac{d\xi}{dr} = -\, \dfrac{1}{r}\left(3\xi + \dfrac{\eta}{\gamma}
	\right) - \dfrac{dp}{dr}\dfrac{\xi}{p+\rho}},
	\end{equation}
	\begin{multline}
	\label{eq12}
	\dfrac{d\eta}{dr} = \xi\left[\omega^{2}r\left(1+\dfrac{\rho}{p}
	\right)e^{\lambda-\chi}-\dfrac{dp}{dr}\dfrac{4}{p}-8\pi(p+
	\rho)\,r\,e^{\lambda}\right.\\\left.+\;\dfrac{r}{p\,(p+\rho)}{\left(\dfrac{dp}{dr}\right)}^{2}\right] +\eta\left[-\,\dfrac{\rho}{p\,(p+\rho)}
	\dfrac{dp}{dr}-4\pi(p+\rho)\,r\,e^{\lambda}\right],
	\end{multline}
where $\omega$ is the eigenfrequency of vibration and $\gamma$ is the 
relativistic adiabatic index, which  plays a significant role in the dynamical 
stability of a star \cite{Chanmugam1977} and is given by
\begin{equation}
\label{13}
        \gamma=\dfrac{dp}{d\rho}(1+\rho/p).
        \end{equation}
This system of coupled first order differential Eqs.\ (\ref{eq11}) and 
(\ref{eq12}) contains singularities at the centre and the surface. To 
solve such system of equations we need two boundary conditions, one at the 
centre as 
$r\rightarrow0$ and other at the surface of the star, i.e. at $r=R$. As 
$r\rightarrow0$, the coefficient of $1/r$ in Eq.\ (\ref{eq11}) must vanish. So 
the condition that must satisfy at the centre or the first boundary condition 
is
\begin{equation}
	\label{eq14}
	3\,\gamma\,\xi+\eta=0.
	\end{equation}
The second boundary condition demands that as $r\rightarrow R$, $\rho
\rightarrow0$, $p\rightarrow0$ and $dp/dr\rightarrow0$. Again as $p
\rightarrow0$, $p/\rho\rightarrow0$ and $\rho/p\rightarrow\infty$. Therefore 
the coefficient of $\rho/p$ in Eq.\ (\ref{eq12}) must vanish as $r\rightarrow R
$. Thus using Eq.\ (\ref{eq9}) in Eq.\ (\ref{eq12}) from this condition it can 
be found that  
\begin{equation}
	\label{eq15}
	\eta=\xi\left[-\,4+\left(1-\dfrac{2M}{R}\right)^{-1}\left(-\,\dfrac{M}{R}-\dfrac{\omega^{2}R^{3}}{M}	
	\right)\right]
	\end{equation}
must be satisfied at the surface of the star. Here $M$ and $R$ are the mass and
radius of the star respectively. These coupled differential Eqs.\ (\ref{eq11}) 
and (\ref{eq12}) along with the boundary conditions Eqs.\ (\ref{eq14}) and 
(\ref{eq15}) form a two point boundary value problem of the Sturm-Liouville type and has 
real eigenvalues $\omega_{0}^{2}<\omega_{1}^{2}< \hdots <\omega_{n}^{2}< 
\hdots$, where $\omega_{n}$ are the eigenfrequencies of oscillations, which 
have $n$ nodes \citep{cox80}. The $n=0$ mode is called as the fundamental or 
$f$-mode. Since $\omega$ is real for $\omega^{2}>0$, hence the solution is 
purely oscillatory in a stable state. But for $\omega^{2}<0$, the frequency 
$\omega$ becomes imaginary and it corresponds to an exponentially growing 
unstable radial oscillations. We solved this couple differential equations by 
using shooting method as described by G. Panotopoulus in \citep{panotopoulos2017}. First we 
have calculated the dimensionless quantity $\bar{\omega}=\omega\, t_{0}$, where 
$t_{0}=1$ ms and then the frequencies are calculated by using the relation :
\begin{equation}\label{eq16}
	\nu=\dfrac{\overline{\omega}}{2\pi}\;\;\mbox{kHz}.
	\end{equation} 
The frequency $\nu$ is allowed to take some particular values called the 
eigenvalues $\nu_{n}$. Each values of $\nu_{n}$ corresponds to a specific 
radial oscillation mode of the star. So it can be inferred that a radial 
oscillation mode of a star is associated with the eigenvalue $\nu_{n}$ and the 
corresponding eigenfunctions $\xi_{n}$ and $\eta_{n}$.


\section{Gravitational wave echo frequencies from strange stars}\label{echo}
The collapse of two massive objects lead to the formation of ultra-compact 
objects. The resulting compact object may be an SS. If such stars fulfill 
some criterion, they can echo GWs. Considering the final object as an SS and 
featuring a photon sphere, the typical echo frequency emitted can be 
calculated. As mentioned earlier, the surface of photon sphere is located at 
$R = 3M$ and the Buchdhal's limit radius is $R = 9 M/4$. Thus to emit GWE, the 
compact stellar object should have compactness $M/R$ larger than $1/3$ and 
smaller than $4/9$. 

To calculate the frequency of GWE, the TOV Eqs.\ (\ref{eq7}) -- (\ref{eq9}) 
together with a EoS are to be solved. For this purpose we consider the EoSs 
described in Sec.\ \ref{eos}. Among these EoSs, those equations are found
to be useful which can mimic a star with larger compactness. When the masses 
of SSs are sufficiently large, the gravitational pull becomes larger and will 
give a more compact configuration. As we have mentioned earlier that only those 
compact stellar objects having compactness within the photon sphere line and 
Buchdahl's limit can emit GWEs, i.e.\ the mass-radius curves should cross the 
photon sphere line, but do not approach the Buchdahl's limit line. To check
this criterion, EoSs given in Eqs.\ (\ref{eq1}) -- (\ref{eq3}) are used in 
this study. In Fig.\ \ref{fig1} the mass-radius curves 
for these EoSs are shown. Also the photon sphere limiting line, the Buchdahl's 
limit line and black hole line are shown in the figure. It is seen that 
the EoSs given by Eq.\ (\ref{eq1}) and Eq.\ (\ref{eq2}) with considered values 
of constants can only give the required compactness and cross the photon sphere 
limit line. Whereas in the case of polytropic EoS, the last stable 
configuration with largest mass is small and hence the compactness.
 
The typical GWE time can be given as the light crossing time from the 
centre of the astrophysical object to the photon sphere \citep{pani2018, 
mannarelli2018},
\begin{equation}
	\label{eq19}
	\tau_{echo}=\int_0^{3M}\!\!\!\! \dfrac{1}{\sqrt{e^{\,\chi(r)}(1-2m(r)/r)}}	
	\; \mathrm{d}r.
	\end{equation} 
The terms $m(r)$ and $\chi(r)$ can be evaluated by solving Eqs.\ (\ref{eq7}) -- 
(\ref{eq9}) for EoS models considered. Using this relation for the 
characteristic echo time, the GWE frequency can be calculated from the 
relation $\omega_{echo}=\pi/\tau_{echo}$ \citep{cardoso1} and the corresponding
repetition frequency of the echo signal can be calculated using the relation 
$\omega_{repetion}=1/(2\,\tau_{echo})$. In general echoes have two natural 
frequencies: the harmonic or resonance frequencies and the quasi-normal mode 
frequencies \citep{abedi2019}. We refer the harmonic or resonance 
frequency as the repetition frequency, as it actually corresponds to the 
repetition frequency of the echo signal. The results obtained for these EoSs 
are discussed in Sec.\ \ref{numerical}.

\section{Numerical results and discussion}\label{numerical}
\begin{figure*}
	\centerline{
	\includegraphics[scale=0.35]{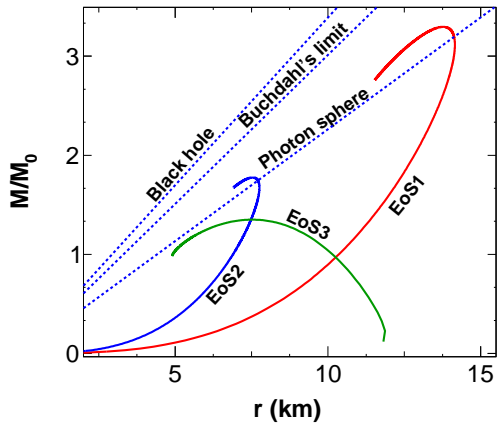}}
	\vspace{-0.3cm}
	\caption{Mass-radius relationships of various SSs for the MIT Bag model 
	 EoS, linear EoS and polytropic EoS showing photon sphere limit, 
     Buchdahl's limit and black hole limit lines. Here EoS1 represents the 
     MIT bag model EoS with $B=(190\,\mbox{MeV})^{4}$, EoS2 is for the 
     linear EoS with $b=0.910$, and the polytropic EoS with $\Gamma=2$ 
     is denoted as EoS3. Although this figure is plotted for one model 
     parameter of each EoS, the patterns of this figure are applicable to all 
     chosen parameters of the corresponding EoSs.} 
	\label{fig1} 
	\end{figure*}

As described above, in this study we have computed radial frequencies of $22$ 
lowest order radial oscillation modes and GWE frequencies of different SSs. For 
the typical SS models considered here, we have chosen maximum masses of SSs, $M
\leq 3.3\,M_{\odot}$ and radii $R\leq 14$ km. The behaviour of such stars is 
different from other compact stars and this can be visualized by 
observing their mass-radius relationships. As already mentioned in the last
section, in Fig.\ \ref{fig1} we have plotted the mass-radius relationships for 
the EoSs described in Sec.\ \ref{eos}. The mass-radius relation for SSs in 
polytropic EoS is following a pattern somewhat different from the other two 
EoSs. It is because, such EoS will give stars with smaller value of 
compactness. For other values of model parameter which are not included in 
Fig.\ \ref{fig1}, the respective compactness and mass-radius behaviour 
can be found in Table \ref{tab:table1}. In this table we have shown the 
values of mass, radius and compactness of SSs obtained for different EoSs. It 
would be appropriate to mention here that according to stability 
analysis of hydrostatic equilibrium configuration of stellar structure under 
polytropic EoS, the equilibrium mass $M$ of such structure varies with respect 
to its central density $\rho_{c}$ as 
$$\frac{d M}{d \rho_{c}} \propto \Gamma - \frac{4}{3}.$$ 
This relation implies that the equilibrium stellar configuration with
$dM/d\rho_{c}\geq 0$ are stable, but those with $dM/d\rho_{c} < 0$ are
unstable \citep{shapiro2004}. In view of this stability condition, all 
three values of $\Gamma = 1.5, 1.67$ and $2$, we have considered in this 
study, should give the stable mass-radius relationship. So, the mass-radius 
curve for the polytropic EoS with $\Gamma = 2$ in Fig.\ \ref{fig1} can be 
viewed as the hydrostatic equilibrium stellar configurations with different 
central densities who satisfy the same polytropic EoS. Thus the highest point 
(along y-axis) on the curve corresponds to maximum mass and density of a 
stable star, whereas the lowest point with maximum radius gives the lowest 
mass and density of a stable star that are governed by polytropic EoS with 
$\Gamma = 2$ in hydrostatic equilibrium. In the following subsections we 
discuss our numerical results on radial oscillation frequencies and GWEs. 
\begin{table*}
\caption{\label{tab:table1} Mass, radius and compactness of SSs for three EoSs 
with different model parameters.}

\begin{ruledtabular}
\begin{tabular}{ccccc}
EoSs & Model & Radius R & Mass M & Compactness \\[-3pt] 
     & Parameter & (in km) & (in $M_{0}$) & (M/R) \\ \hline
\multirow{3}{*}{MIT Bag model} 
     & $B=(190\, \mbox{MeV})^{4}$ & 13.766 & 3.295 & 0.3540\\
     & $B=(217\, \mbox{MeV})^{4}$ & 10.630 & 2.544 & 0.3540\\
     & $B=(243\, \mbox{MeV})^{4}$ &  8.456 & 2.024 & 0.3540\\
\multirow{3}{*}{Linear EoS} 
     & $b=0.910$ & 7.535 & 1.775 & 0.3484\\
     & $b=0.918$ & 7.816 & 1.844 & 0.3489\\ 
     & $b=0.926$ & 8.128 & 1.920 & 0.3494 \\ 
\multirow{3}{*}{Polytropic EoS} 
     & $\Gamma=1.50$ & 11.200 & 0.814 & 0.1081\\
     & $\Gamma=1.67$ &  7.980 & 0.964 & 0.1790\\ 
     & $\Gamma=2.00$ &  7.500 & 1.350 & 0.2600 \\ 
\end{tabular}
\end{ruledtabular}
\end{table*}
\subsection{Radial oscillation frequencies of SSs}
\begin{figure*}
        \centerline{
        \includegraphics[scale = 0.27]{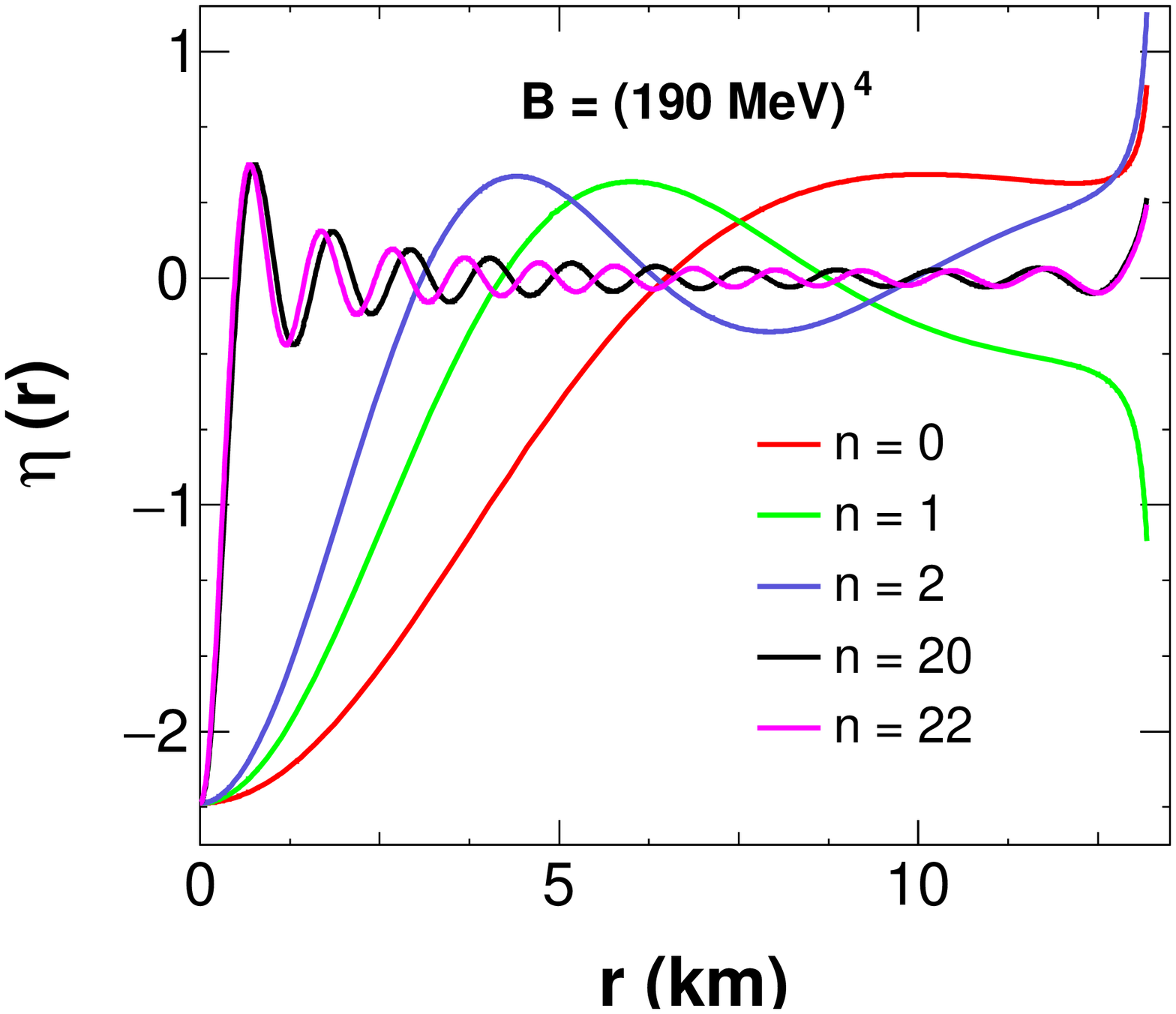}\hspace{0.2cm}
        \includegraphics[scale = 0.27]{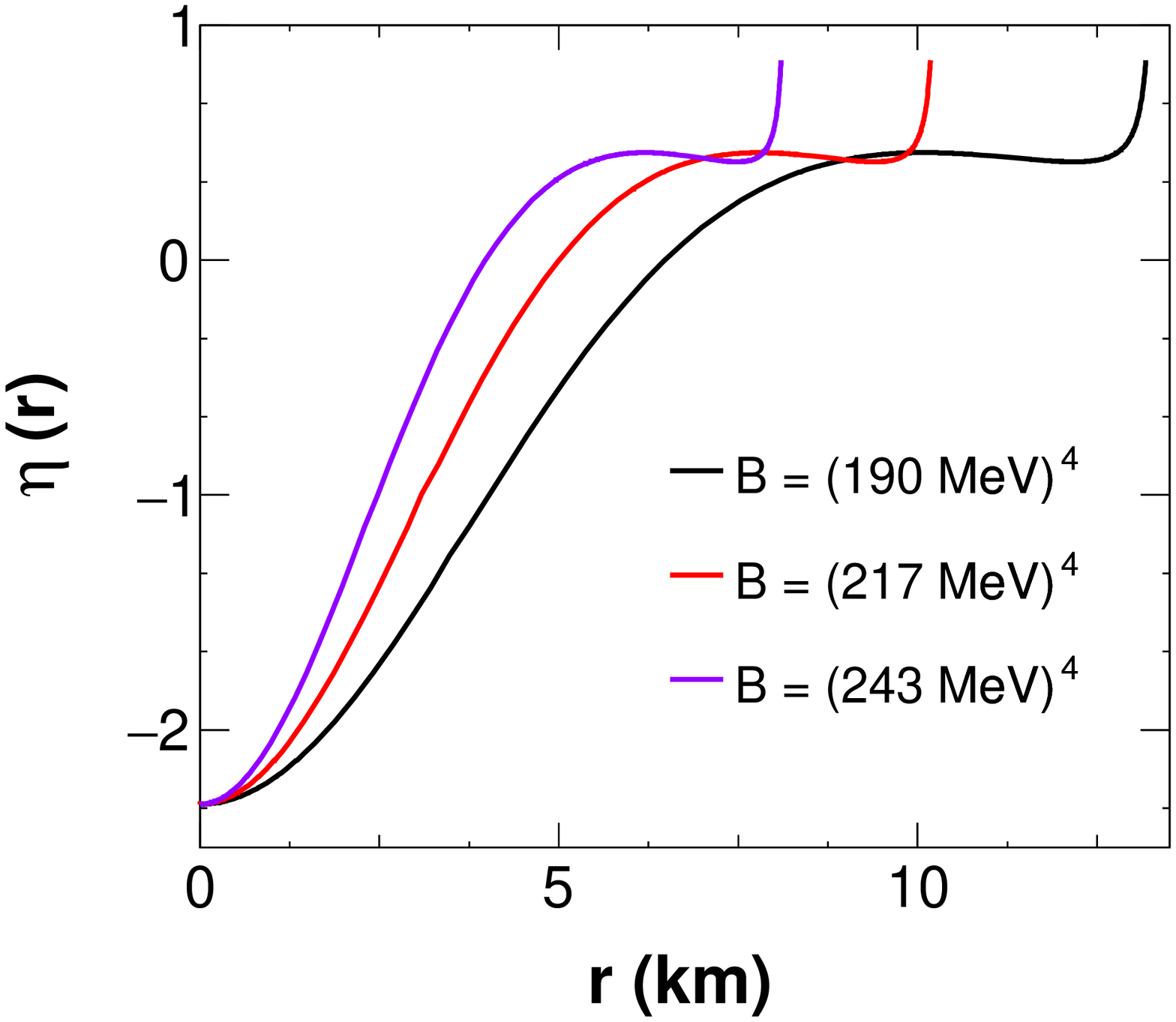}\hspace{0.2cm}
        \includegraphics[scale =0.27]{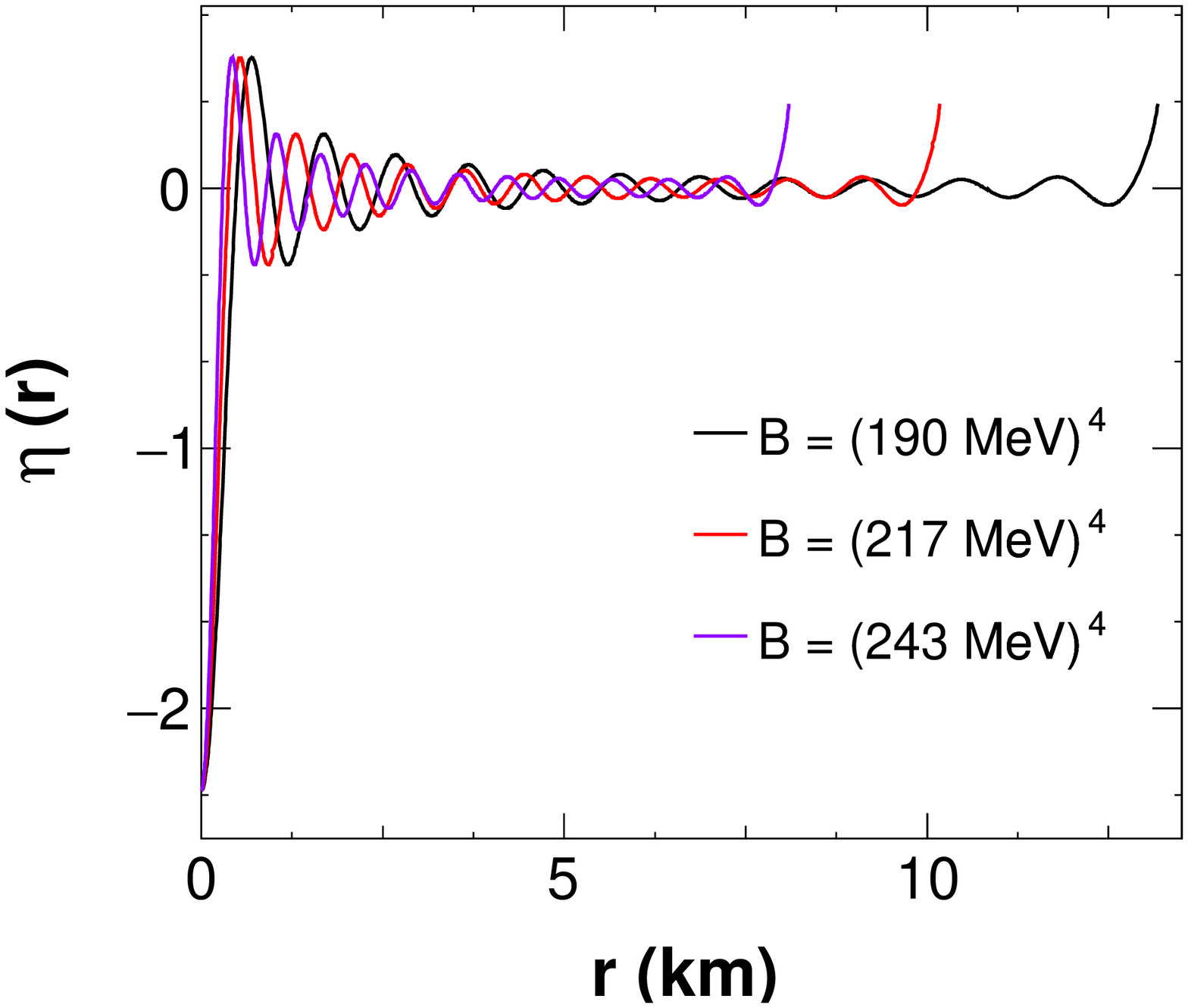}}
        \vspace{-0.3cm}
        \caption{Behaviours of pressure perturbation parameter $\eta (r)$ as a
        function of distance from the centre of SS for different modes of
        oscillations obtained from the numerical solution of Eq.\ (\ref{eq12})
        using the MIT Bag Model EoS. Left plot is with Bag constant $B={(190\,  
        \mbox{MeV})}^4$ for both low order modes $n = 0, 1,2$ and highly
        excited modes $n = 20, 22$. The other two plots are with three Bag
        constants, viz., $B={(190\, \mbox{MeV})}^4$, ${(217\, \mbox{MeV})}^4$
        and ${(243\, \mbox{MeV})}^4 $ for $f$-mode (middle panel) and
        $p_{22}$-mode (right panel) of oscillations respectively.}
        \label{fig5}
        \end{figure*}

\begin{figure*}
        \centerline{
        \includegraphics[scale = 0.27]{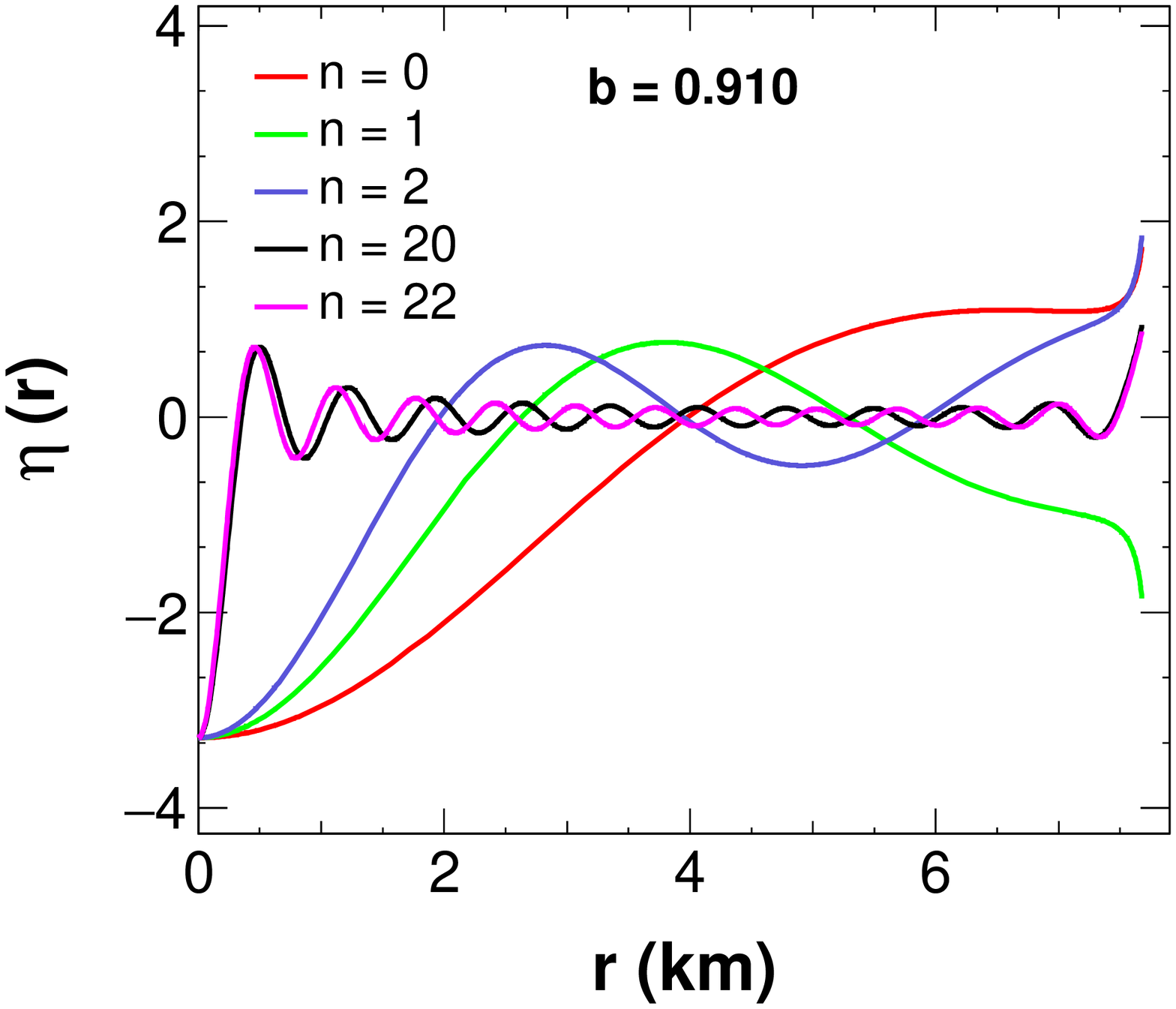}\hspace{0.2cm}
        \includegraphics[scale = 0.27]{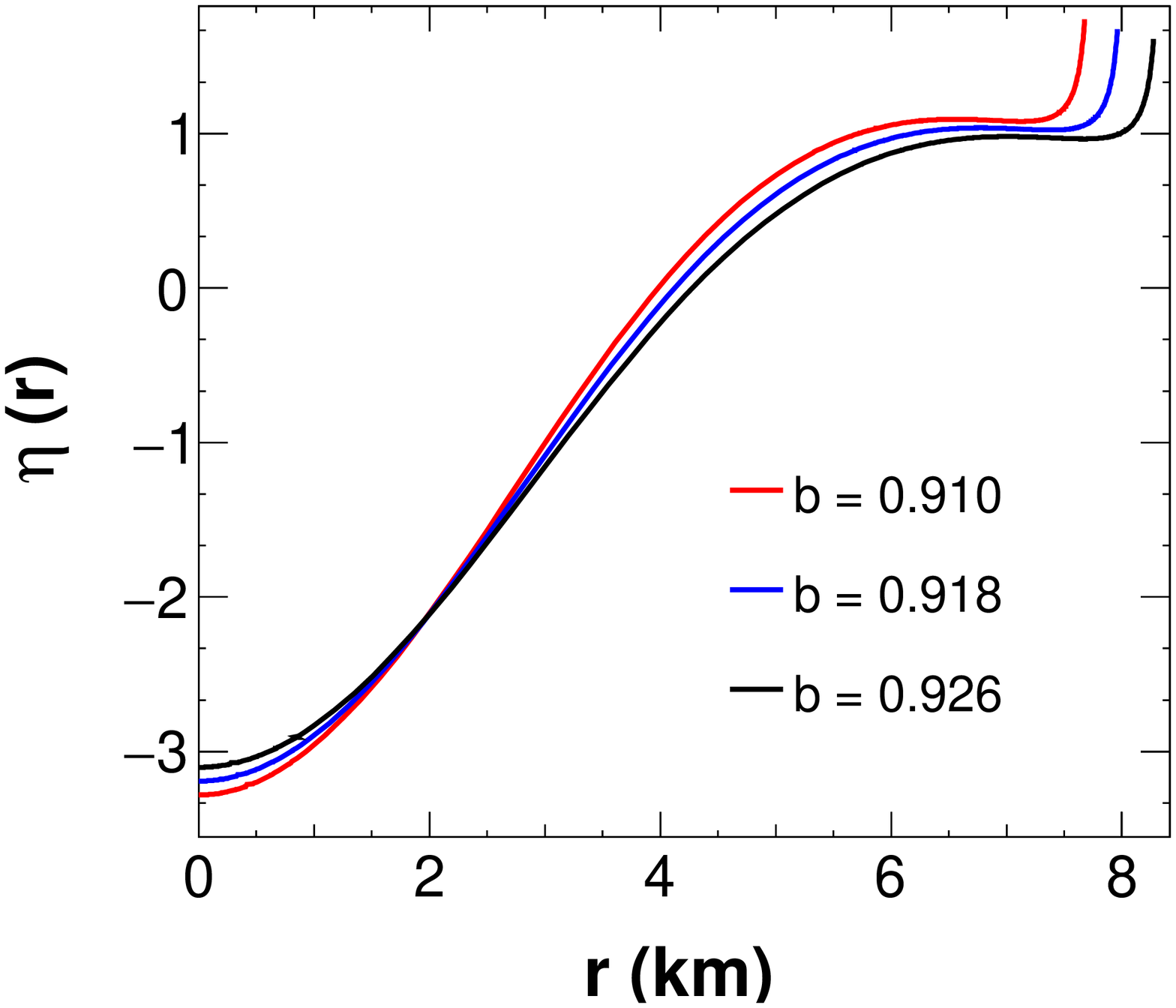}\hspace{0.2cm}
        \includegraphics[scale = 0.27]{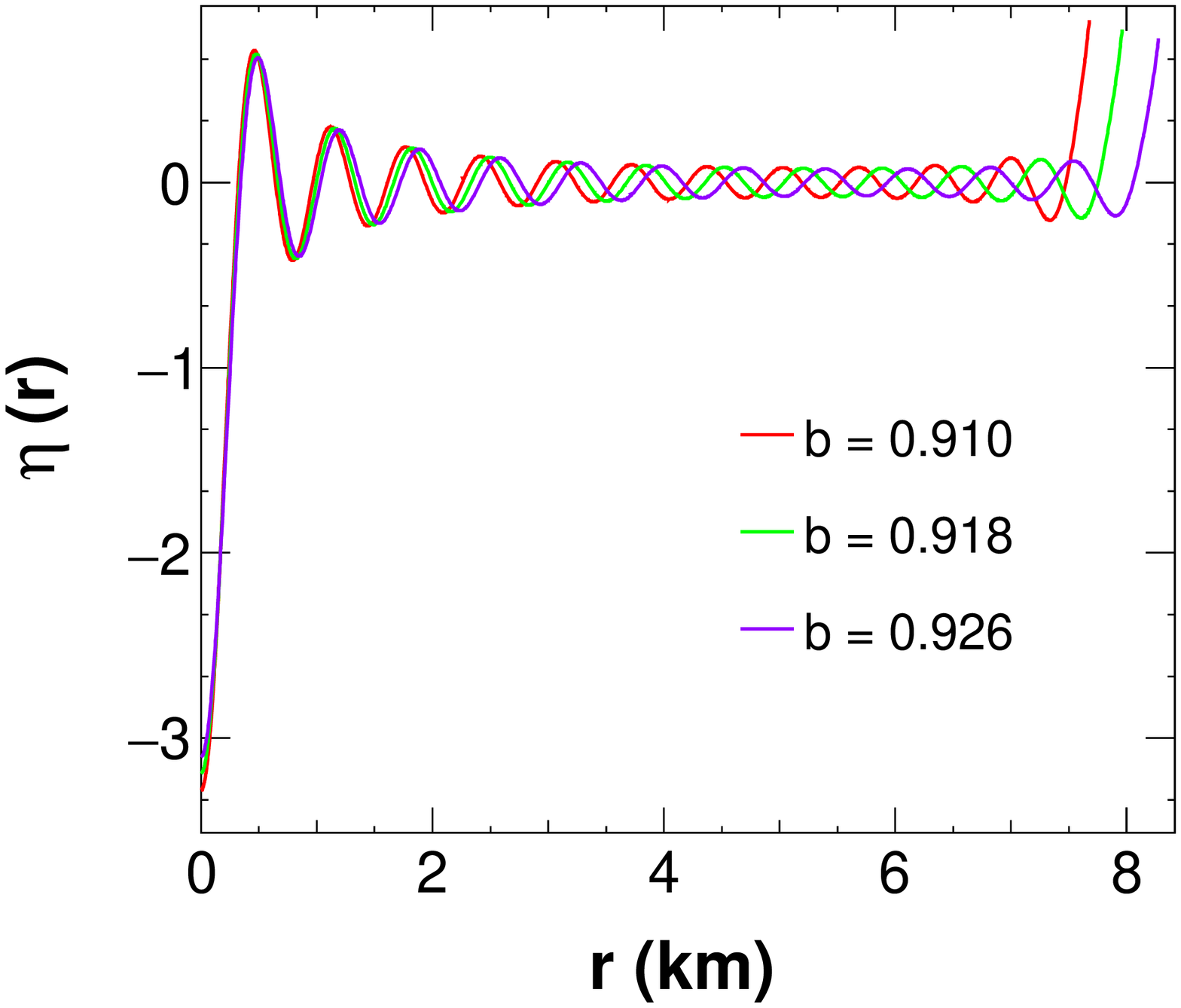}}
        \vspace{-0.3cm}
        \caption{Behaviours of pressure perturbation parameter $\eta (r)$ as a
        function of distance from the centre of SS for different modes of
        oscillations obtained from the numerical solution of Eq.\ (\ref{eq12})
        using the linear EoS. The first plot is with linear constant $b=0.910$
        for both low order modes $n = 0, 1,2$ and highly excited modes $n = 20,
        22$, and the last two plots are for $f$-mode and $p_{22}$-mode of
        oscillations respectively with linear constants $b=0.910$, $0.918$
        and $0.926$.}
       \label{fig6}
       \end{figure*}

\begin{figure*}
        \centerline{
        \includegraphics[scale = 0.27]{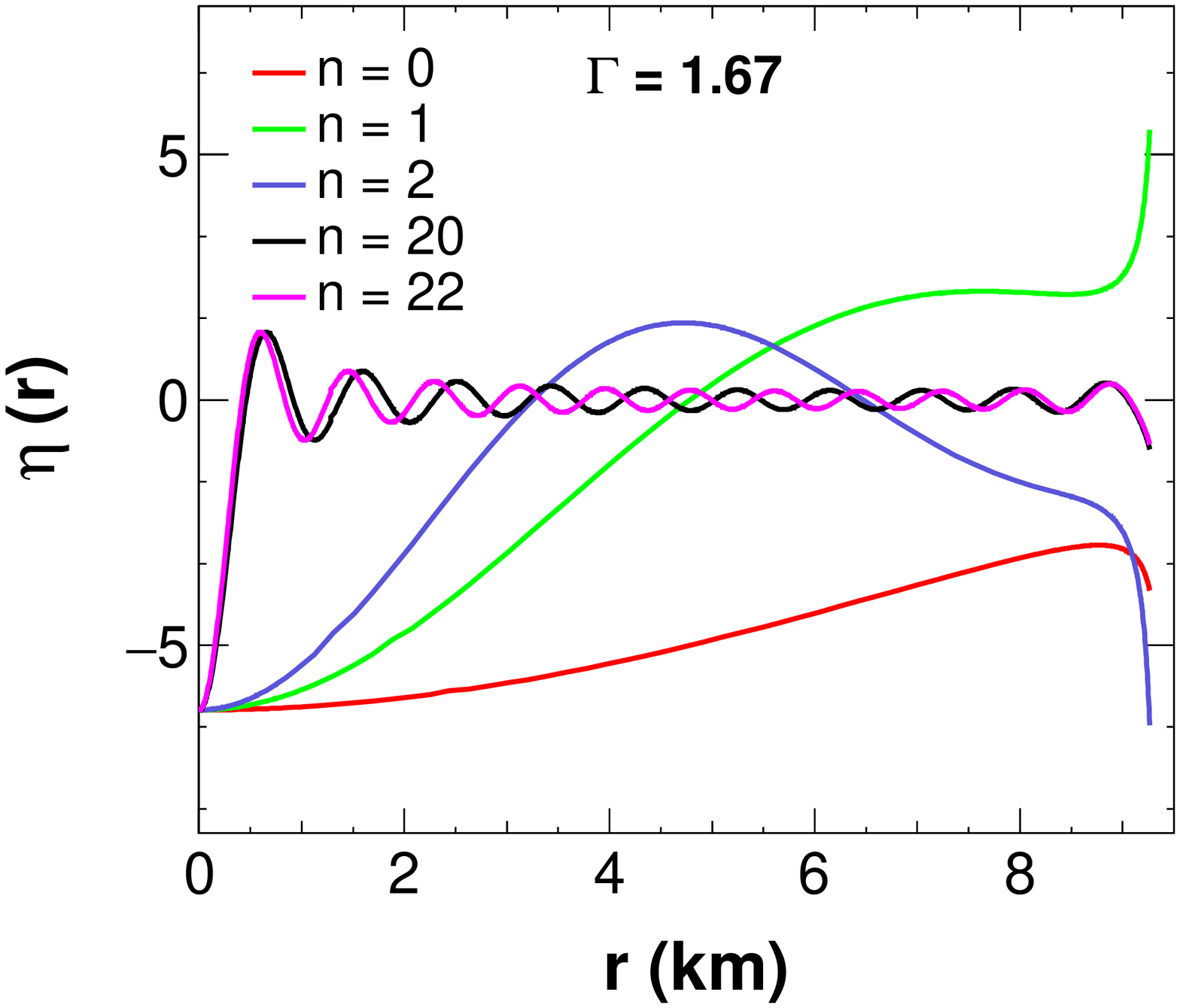}\hspace{0.2cm}
        \includegraphics[scale = 0.27]{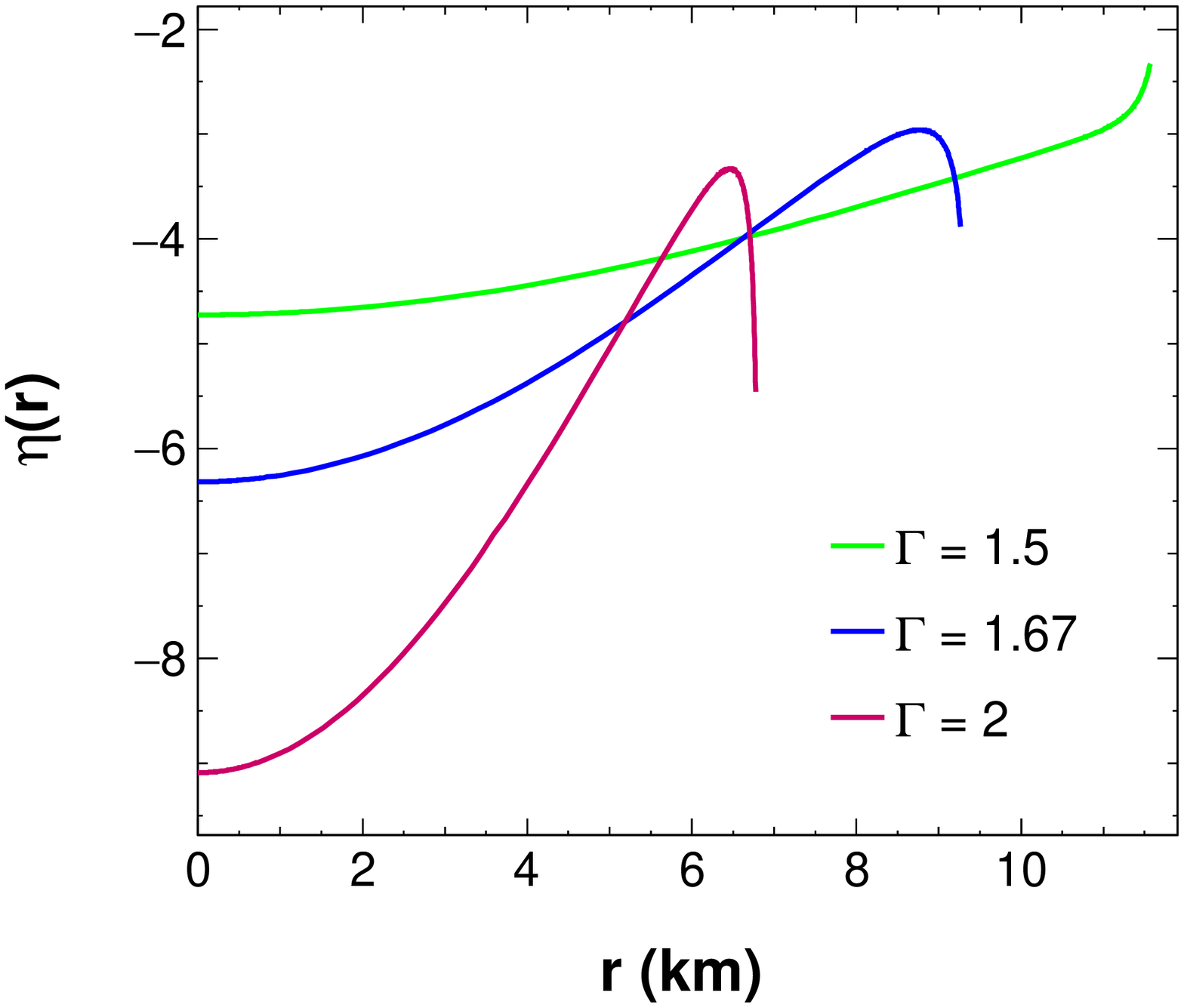}\hspace{0.2cm}
        \includegraphics[scale = 0.27]{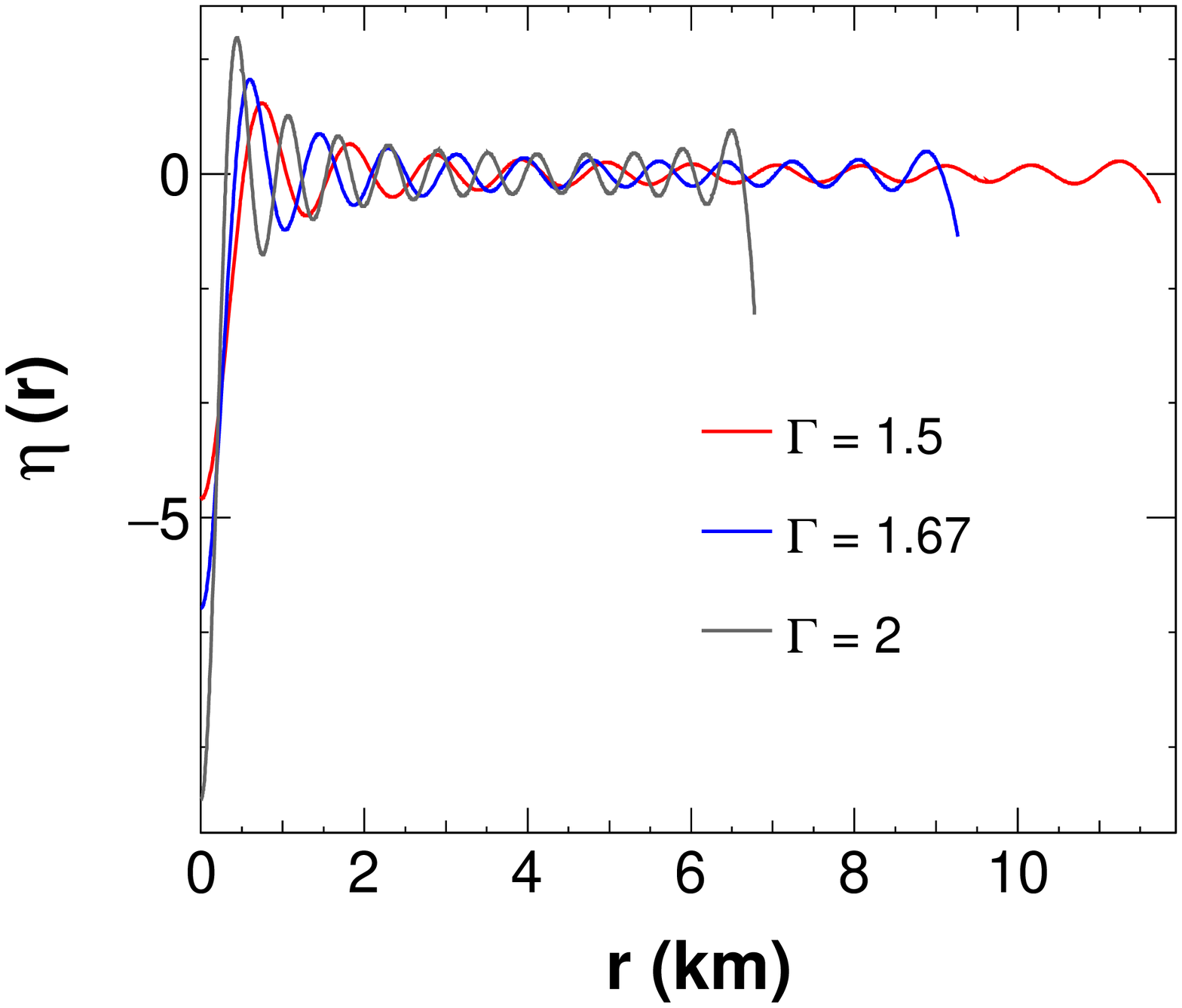}}
        \vspace{-0.3cm}
        \caption{Behaviours of pressure perturbation parameter $\eta (r)$ as
        function of radial distance of SS for low order oscillation modes
        $n = 0, 1, 2$ and highly excited modes $n = 20, 22$ obtained from the
        numerical solution of Eq.\ (\ref{eq12}) using polytropic EoS for $      
        \Gamma=1.67$ [left plot]. Same for $f$-mode [middle
        plot] and $p_{22}$-mode [right plot] with different values of
        polytropic exponent $\Gamma$.}
        \label{fig7}
        \end{figure*}

As mentioned in Sec.\ \ref{radial}, the radial and pressure perturbation
Eqs.\ (\ref{eq11}) and (\ref{eq12}) are the coupled differential equations of
Sturm-Liouville type whose eigenvalue solutions give the various modes of
oscillations, e.g.\ for $n=0$, we get the fundamental mode, for $n=1$  the
first overtone or the pressure $p_{1}$ mode, for $n=2$ the second overtone or
the $p_{2}$ mode and so on. The displacement and pressure perturbation
variables $\xi(r)$ and $\eta(r)$ obtained respectively from the solutions
of these equations are plotted for various modes, mainly for lower order
($n=0,1,2$) modes and for higher order ($n=20,22$) modes, against the distance
$r$ (in km) from the centre of SS. Figs.\ \ref{fig5}, \ref{fig6} and
\ref{fig7} correspond to the plots of the pressure perturbations with respect
to distance from the centre of the star for various modes oscillations 
for MIT
Bag model EoS, linear EoS and polytropic EoS respectively. For these models,
the pressure perturbations $\eta_{n}(r)$ are larger near the centre and 
near the surface of the star. However $\eta_{n}(r)$ is slightly smaller near
the surface of the stars with these EoSs. The values of pressure perturbations
are found to vary from model to model. Also, for a certain model with different
model parameters significant variations are noticed in pressure perturbations.

For the MIT Bag model with $B={(190\,\mbox{MeV})}^4$, the variations of
$\eta_n(r)$ with respect to $r$ for all said modes are shown in the left plot
of Fig.\ \ref{fig5}. The dependence of $\eta_n(r)$ on Bag constant $B$ is shown
in other two plots of this figure. It is seen that for both lower (middle plot
of Fig.\ \ref{fig5}) and higher (right plot of Fig.\ \ref{fig5}) order modes,
the pressure perturbations varies noticeably with $B$. As different values of
Bag constant of MIT Bag model will represent different stars, so the pressure
perturbation along the radius of the star will be different for different Bag
constants. The Bag model corresponding to $B={(190\,\mbox{MeV})}^4$ gives
a star with the maximum radius than other two values of $B$. The values of
pressure perturbations $\eta_n(r)$ for MIT Bag model with
$B={(190\,\mbox{MeV})}^4$, ${(217\,\mbox{MeV})}^4$ and
${(243\,\mbox{MeV})}^4$ are almost same near the centre of the star. But at
the surface of the star values of $\eta_n(r)$ are found to be different
slightly as the radius of the star is different with these three different
values of $B$.

As mentioned above, in Fig.\ \ref{fig6} pressure perturbations for linear EoS
are plotted against the radial distance of the star. In the left plot of this
figure the perturbations obtained by using the linear constant $b=0.910$ are
shown for five modes, three of them are lowest order modes and two are highest
order modes as in the earlier case. The perturbations are larger near the
centre and surface of the star. In the middle and right plots of
Fig.\ \ref{fig6} lower and higher order modes are drawn respectively for three
values of linear constants, viz., $b=0.910$, $0.918$ and $0.926$. It is seen
that for the linear EoS with these three constants all modes of perturbations
in pressure are almost same from centre to near the surface of the star. The
highest value of $b$, i.e. $0.926$ considered here corresponds to the star with
maximum radius and the lowest value $b=0.910$ gives star of minimum radius.

For polytropic EoS, the variation of pressure perturbations $\eta_n(r)$ along
the radial distance $r$ of the star is shown in Fig.\ \ref{fig7} as mentioned
earlier. Similar to the cases of linear EoS and MIT Bag model EoS, the
perturbations at the surface of the star are nearly equal (but not exactly) to
the values near the centre of the star. In the left plot of the figure the
perturbed pressures are shown for three lower order modes and two higher order
modes as in the cases of other EoSs. The middle and the right plots are to show
the variation of $\eta (r)$ with $\Gamma$. For all these modes the
perturbations are found to follow almost the same rule: larger near the 
centre and near the surface of the star. However, there is a small 
exception for the case of $n=0$ mode. In this case the amplitude of pressure 
perturbation shows the decreasing trend towards the surface of the star, 
especially for smaller values of $\Gamma$. For the three values of $\Gamma$ 
considered here, the variation in $\eta_n(r)$ with radius is following the 
same pattern with the expecption as mentioned above and the largest 
$\Gamma$ gives the star with smallest radius.

\begin{figure*}
        \centerline{
        \includegraphics[scale = 0.27]{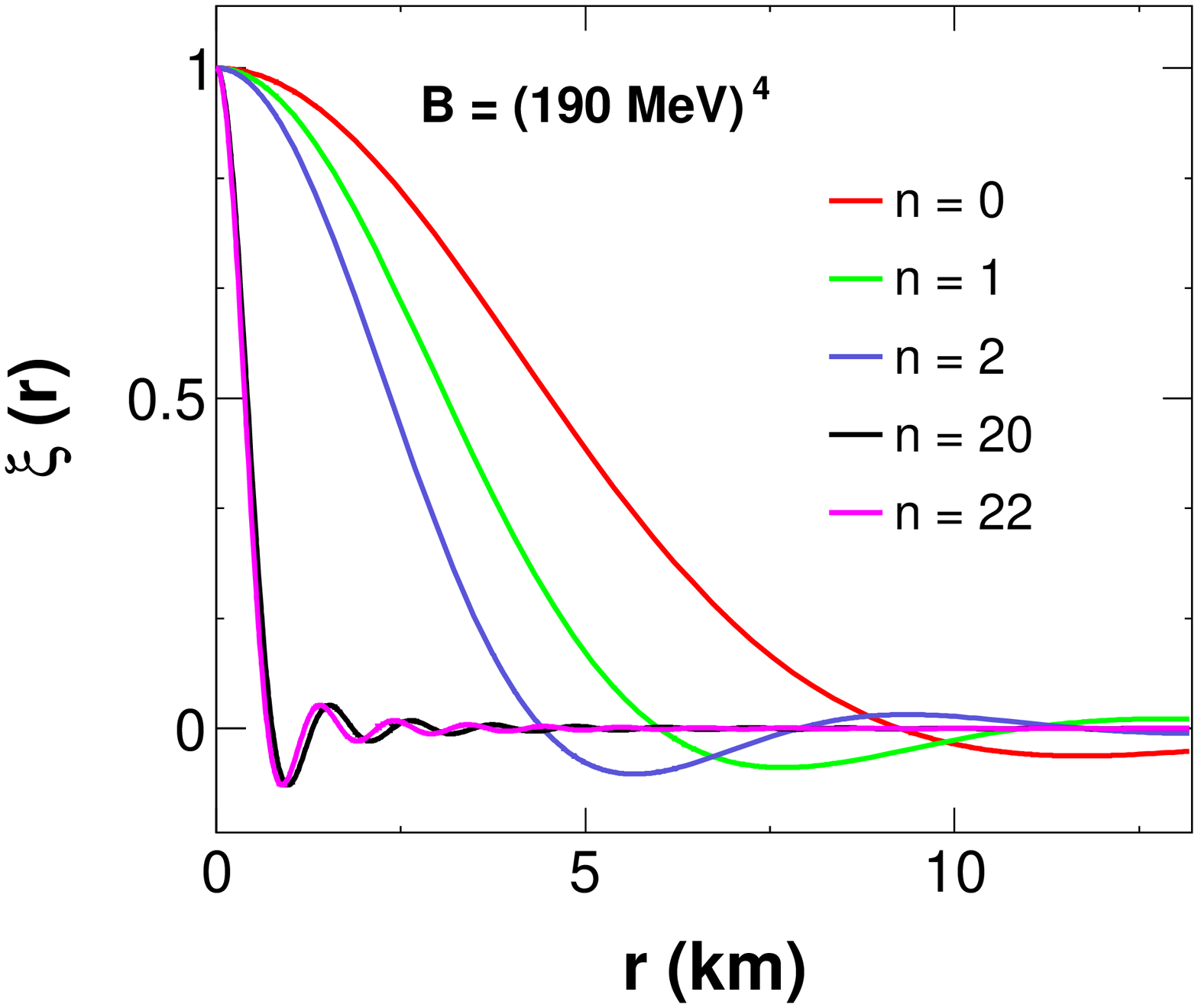}\hspace{0.2cm}
        \includegraphics[scale = 0.27]{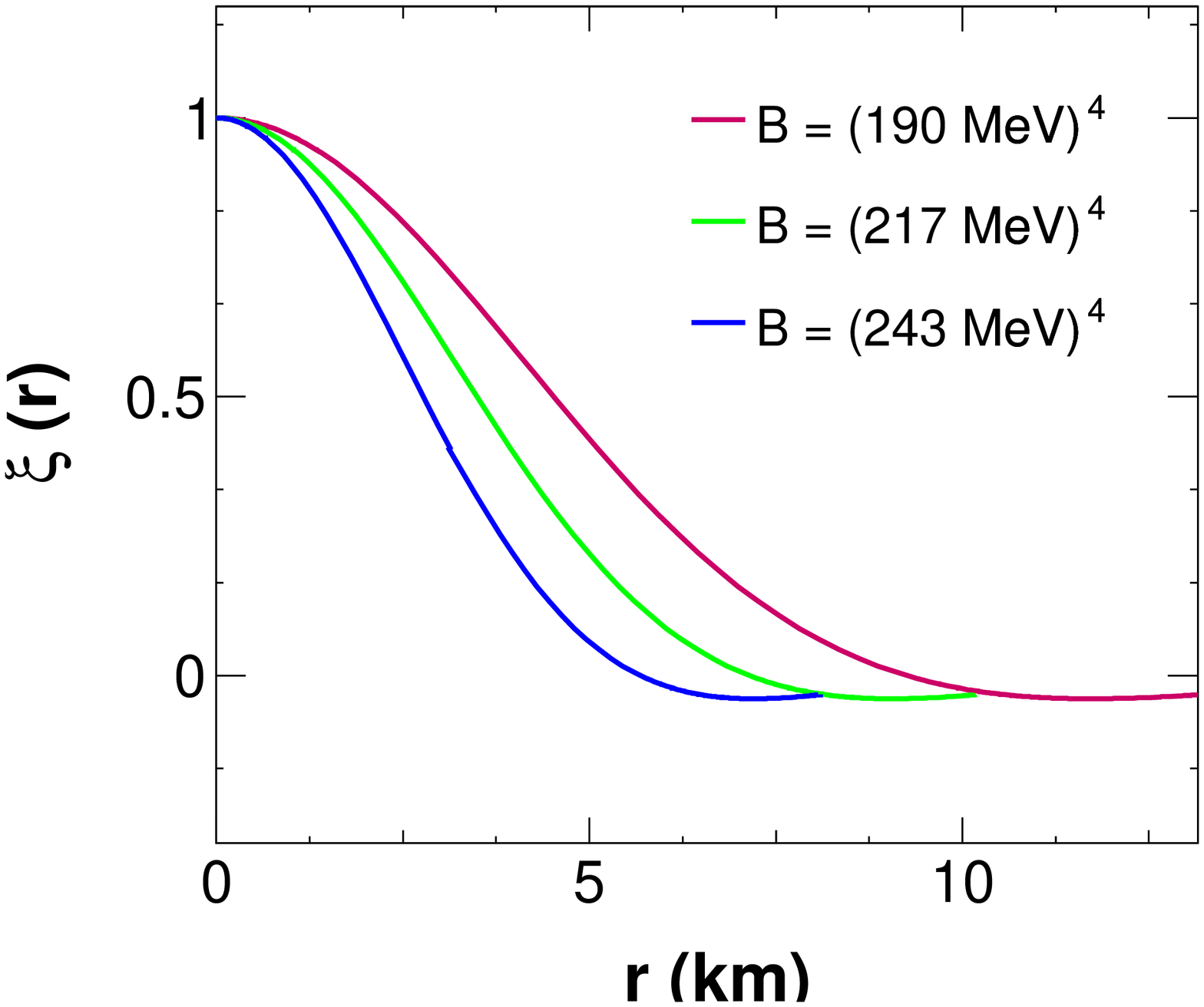}\hspace{0.2cm}
        \includegraphics[scale = 0.27]{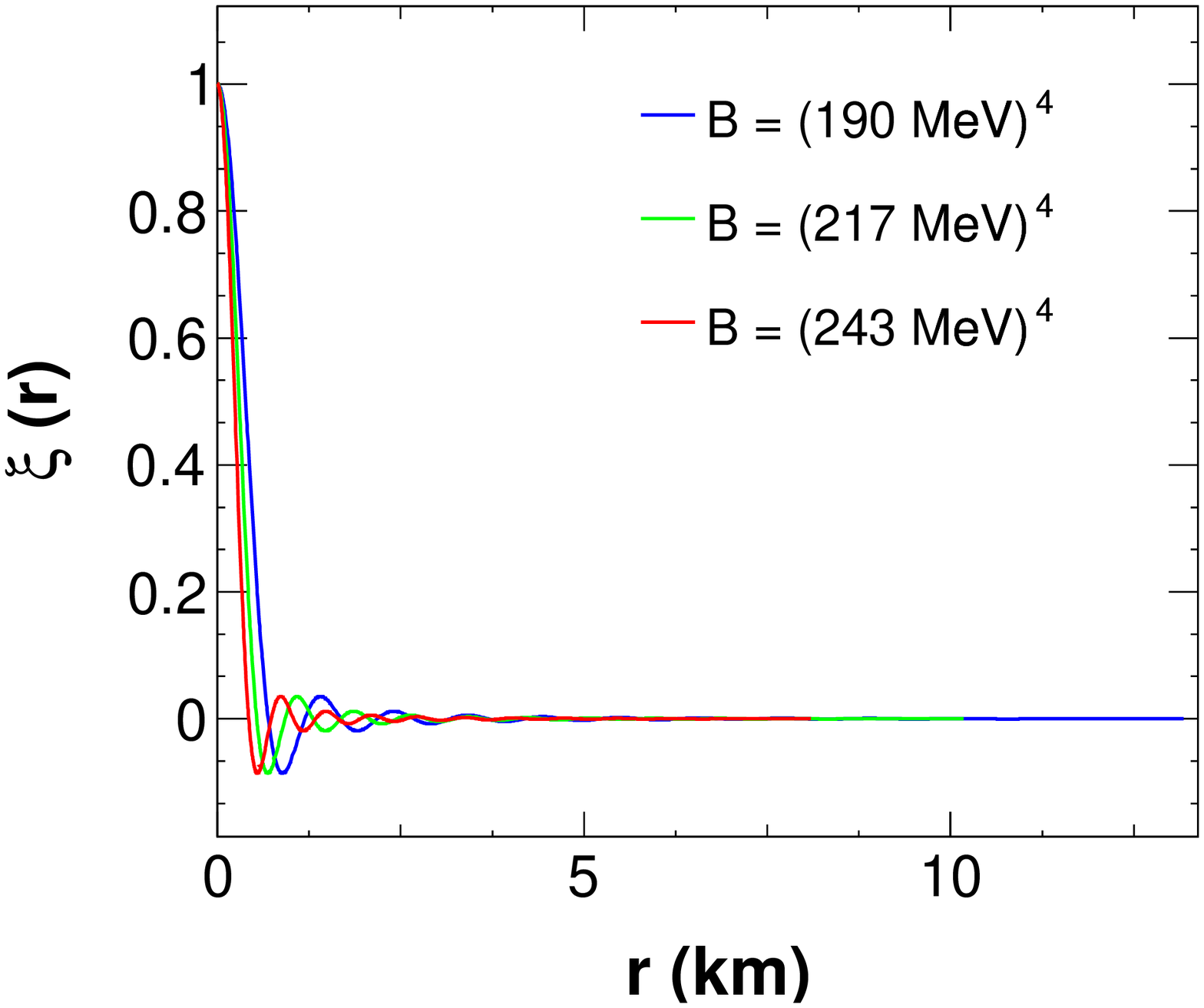}}
        \vspace{-0.3cm}
        \caption{Behaviours of radial perturbation parameter $\xi (r)$ as a
function of distance from the centre of SS for different modes of 
oscillations
obtained from the numerical solution of Eq.\ (\ref{eq11}) using the MIT Bag
Model EoS. First plot is with Bag constant $B={(190\, \mbox{MeV})}^4$ for both
low order modes $n = 0, 1,2$ and highly excited modes $n = 20, 22$, and the
other two plots are for $f$-mode and $p_{22}$-mode of oscillations
respectively with three values of $B={(190\, \mbox{MeV})}^4$, ${(217\,
\mbox{MeV})}^4$ and ${(243\, \mbox{MeV})}^4$.}
       \label{fig8}
       \end{figure*}

\begin{figure*}
        \centerline{
        \includegraphics[scale = 0.27]{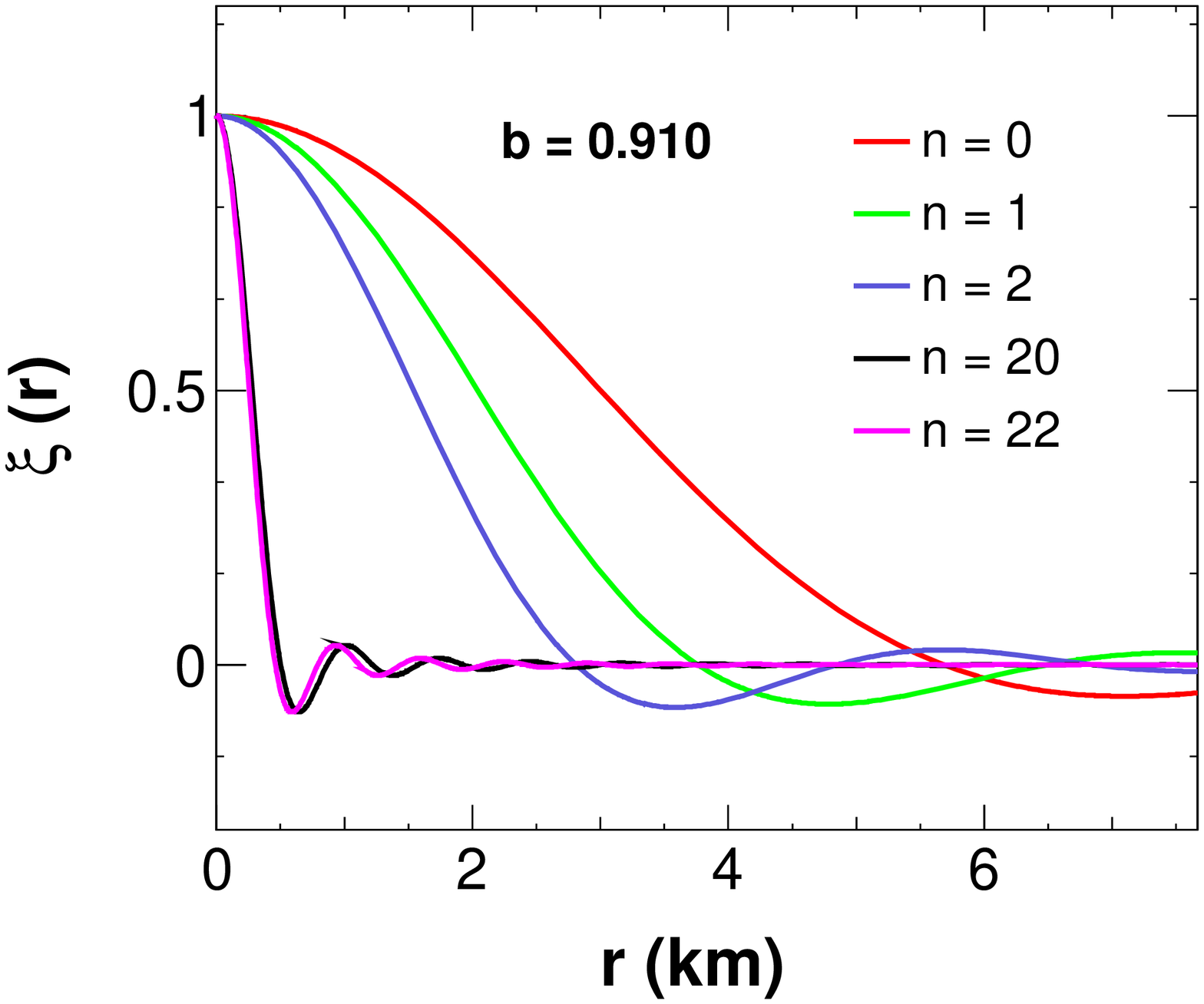}\hspace{0.2cm}
        \includegraphics[scale = 0.27]{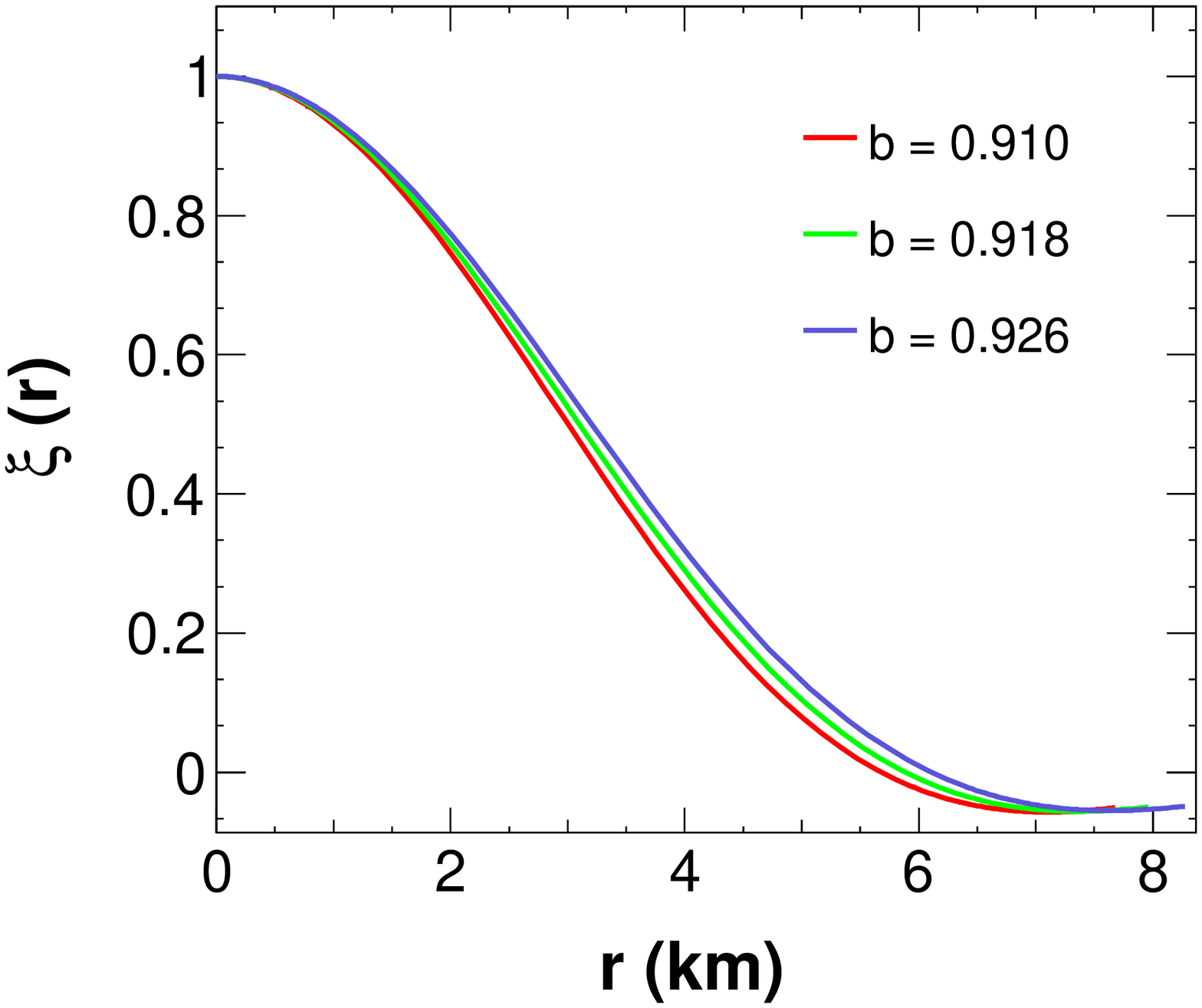}\hspace{0.2cm}
        \includegraphics[scale = 0.27]{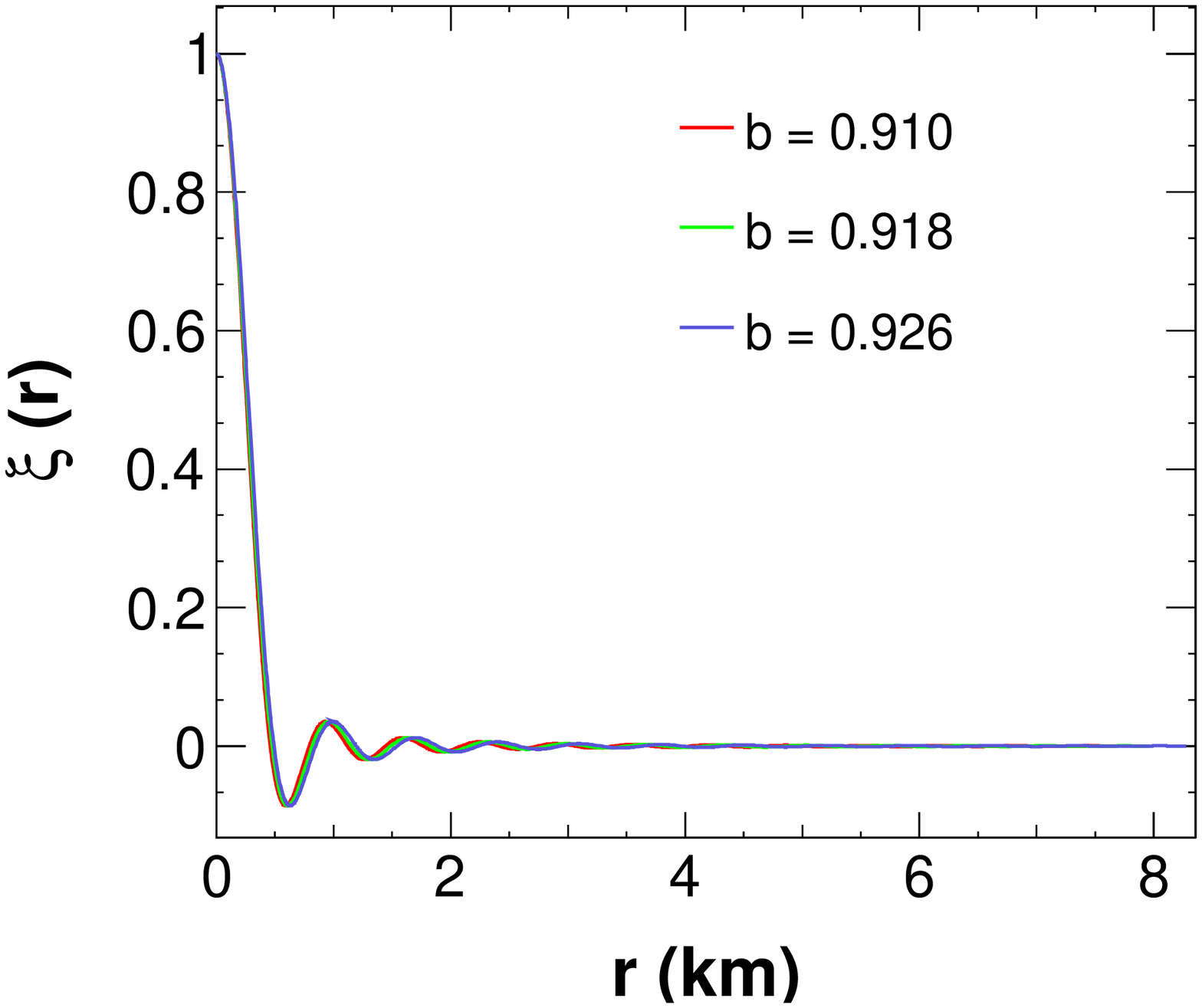}}
        \vspace{-0.3cm}
        \caption{Behaviours of radial perturbation parameter $\xi (r)$ as a
function of distance from the centre of SS for different modes of oscillations
obtained from the numerical solution of Eq.\ (\ref{eq11}) using the linear
EoS. The first plot is with $b=0.910$ for both low order modes $n = 0, 1,2$ and
highly excited modes $n = 20, 22$, and the last two plots are with $b=0.910$,
$0.918$ and $0.926$ for $f$-mode and $p_{22}$-mode of oscillations
respectively.}
        \label{fig9}
        \end{figure*}

\begin{figure*}
        \centerline{
        \includegraphics[scale =0.27]{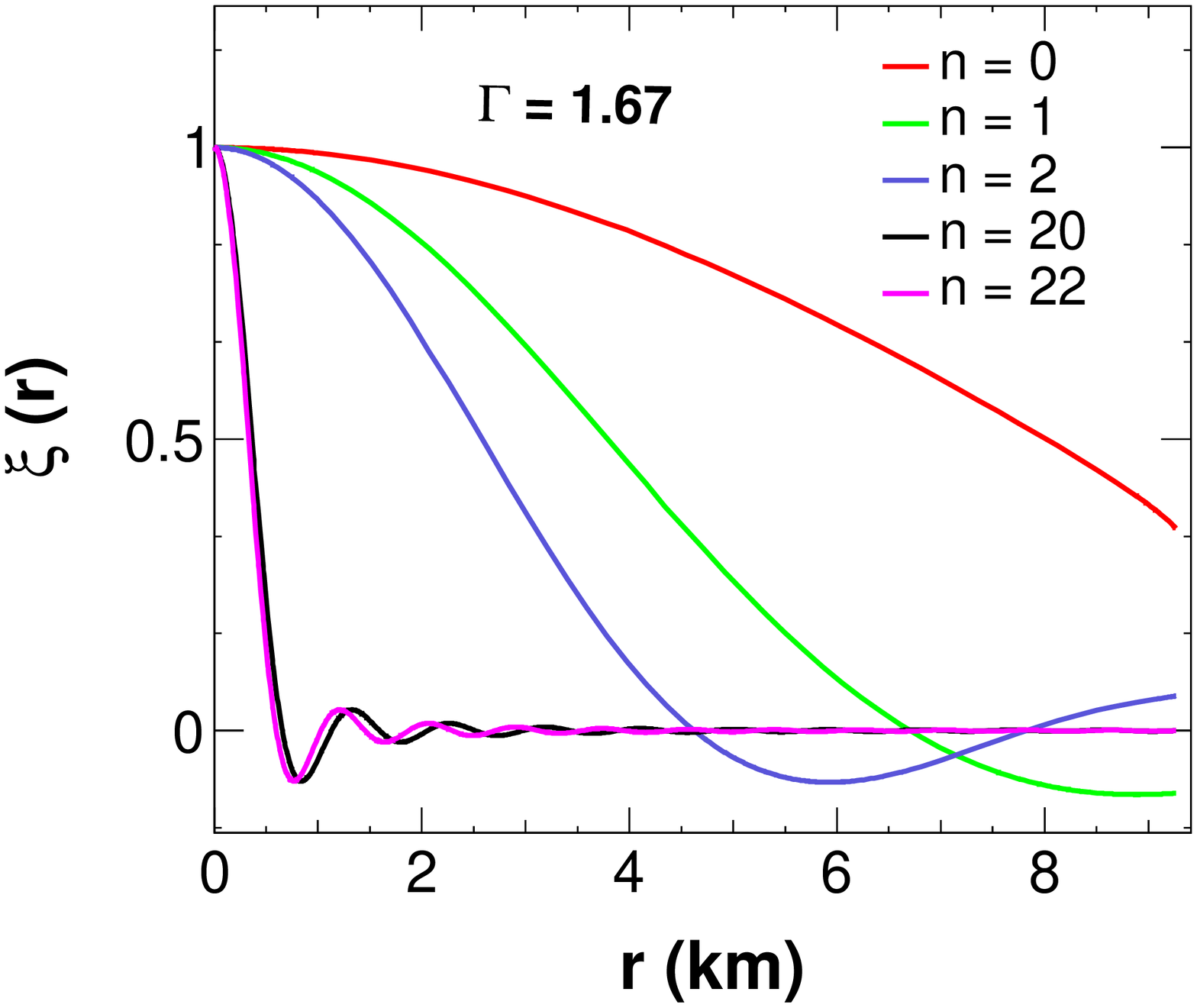}\hspace{0.2cm}
        \includegraphics[scale =0.27]{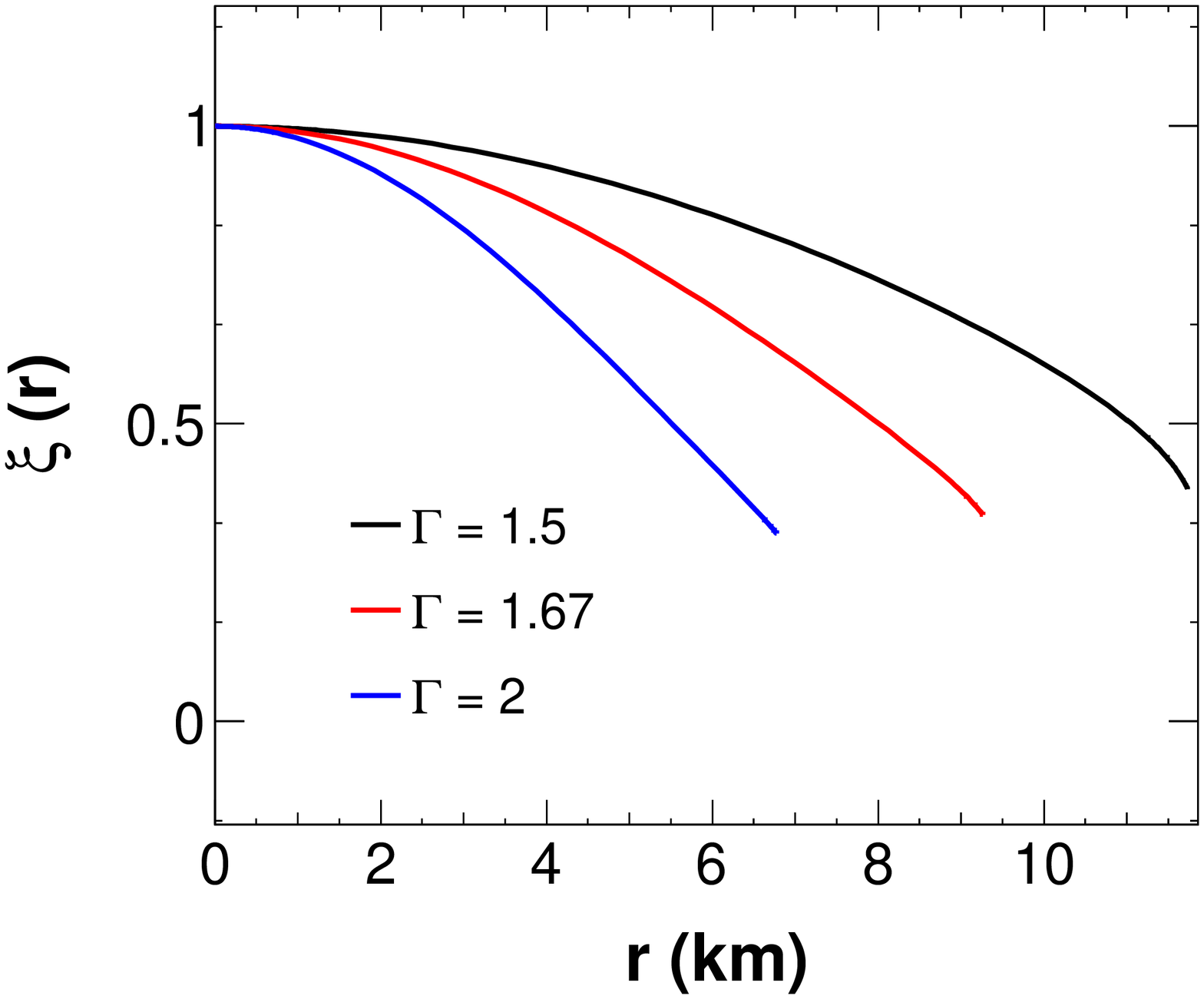}\hspace{0.2cm}
        \includegraphics[scale =0.27]{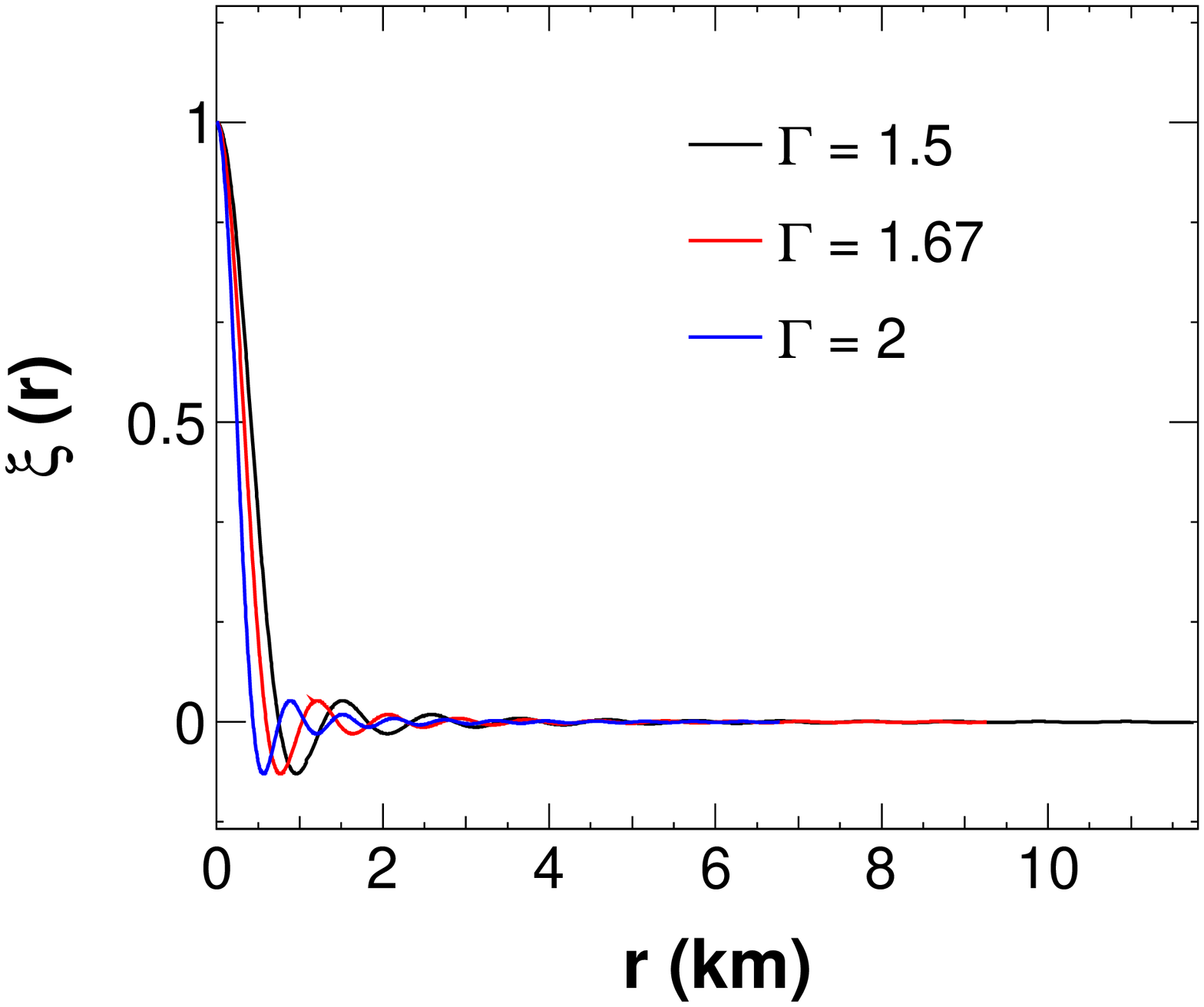}}
        \vspace{-0.3cm}
        \caption{Behaviours of radial perturbation parameter $\eta (r)$ as a
        function of radial distance of SS for low order oscillation modes
        $n = 0, 1, 2$ and highly excited modes $n = 20, 22$ obtained from the
        numerical solution of Eq.\ (\ref{eq11}) using polytropic EoS for $      
        \Gamma=1.67$ [left plot]. Same for $f$-mode [middle
    plot] and $p_{22}$-mode [right plot] with various polytropic exponent $
    \Gamma$ values.}
    \label{fig10}
    \end{figure*}

The radial perturbations of SSs with respect to radial distance obtained
from the numerical solution of Eq.\ (\ref{eq11}) using different EoSs are
shown in Fig.\ \ref{fig8}, \ref{fig9} and \ref{fig10}. Unlike the variation of
pressure perturbations, which are larger near the centre of the star and also
near the surface of the star, radial perturbations are maximum near the
centre of the star only. That is the radial perturbations gradually decrease
with along the radial distance and become minimum near the surface of the star.
For MIT Bag model with $B={(190\, \mbox{MeV})}^4$, the variation of $\xi_n(r)$
with $r$ is shown in the left plot of Fig.\ \ref{fig8}. The other two plots of
this figure are to show how these radial perturbations depend on the Bag
constants. In the middle plot the dependence of low order oscillation modes on
$B$ is shown and in the right plot it is shown for higher order modes. The
low value of $B$ corresponds the maximum radius as clear from the middle and
right plots of the figure.

Same study is made for the linear EoS with three different linear constants as
earlier, which is represented in Fig.\ \ref{fig9}. It is already seen that
stars given by different linear constants have slightly varying radii. The
variation of $\xi_n (r)$ for different $b$ is found to be smaller than the
variation of $\xi_n(r)$ for stars (i.e. for different $B$) represented by MIT
Bag models. In the left plot different modes of $\xi(r)$ is plotted against
radial distance of the star for $b=0.910$. The middle and the right plots of
this figure correspond to the variation of $\xi_n(r)$ with different values of
$b$ for low order modes and higher order modes respectively. As in the case of
pressure perturbations, the radial perturbations also show that for the star
with $b=0.926$ will have larger radius (but slightly) and hence has maximum
perturbation compared to the other two values of $b$.

For polytropic EoS, with polytropic index $\Gamma=1.5$, $1.67$ and $2$ the 
radial perturbations $\xi_n(r)$ are plotted for different oscillation
modes of the star in Fig.\ \ref{fig10}. As in the case of other EoSs, for this
model, $\xi(r)$ is maximum near the centre of the star and minimum near the
surface. The middle and right plot of this figure show the variation of
$\xi_n(r)$ with different $\Gamma$ values. We have seen that $\xi_n(r)$
vary significantly with $\Gamma$, specially in low order modes.

\begin{figure*}
        \centerline{
        \includegraphics[scale = 0.3]{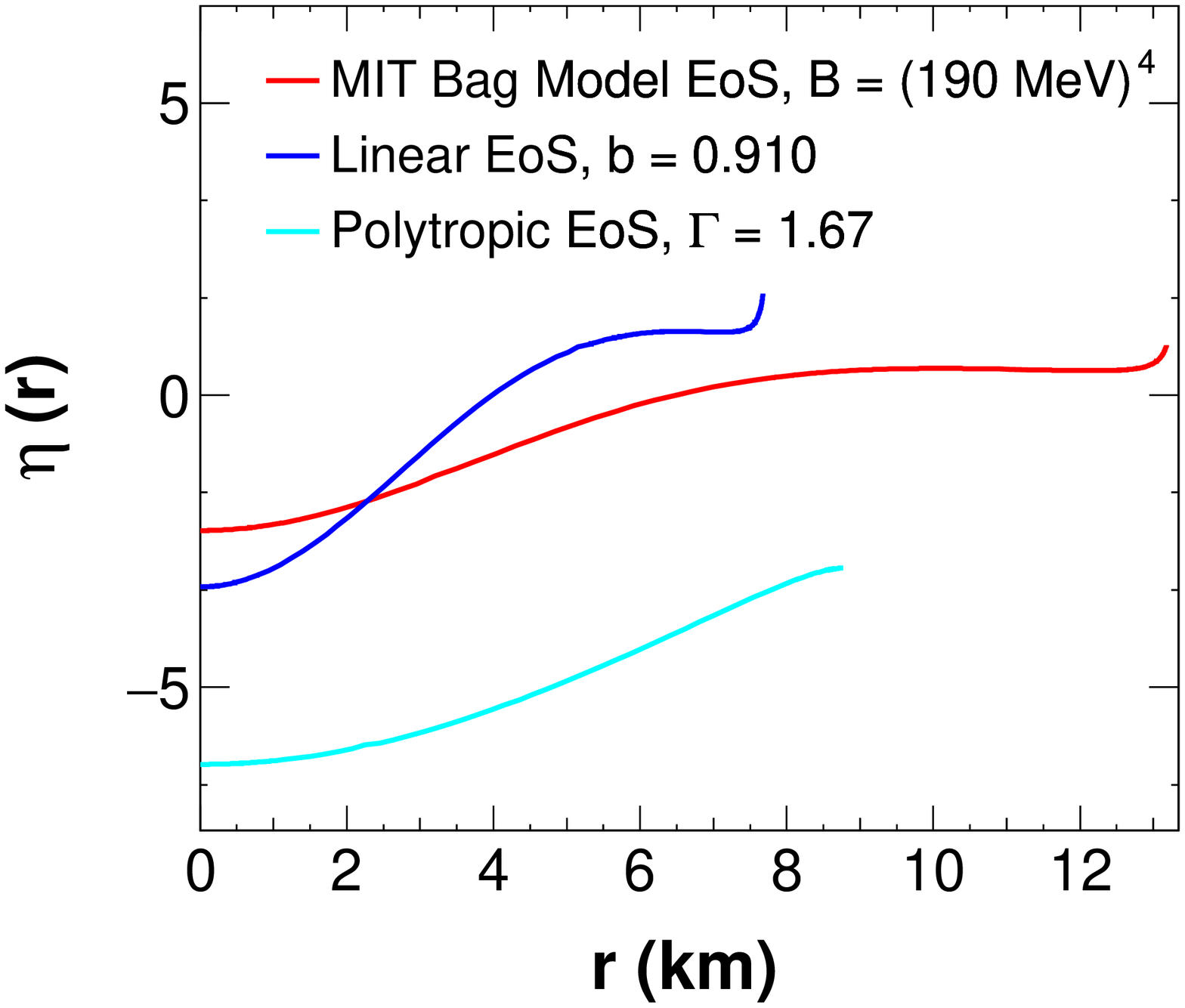}\hspace{0.5cm}
        \includegraphics[scale = 0.3]{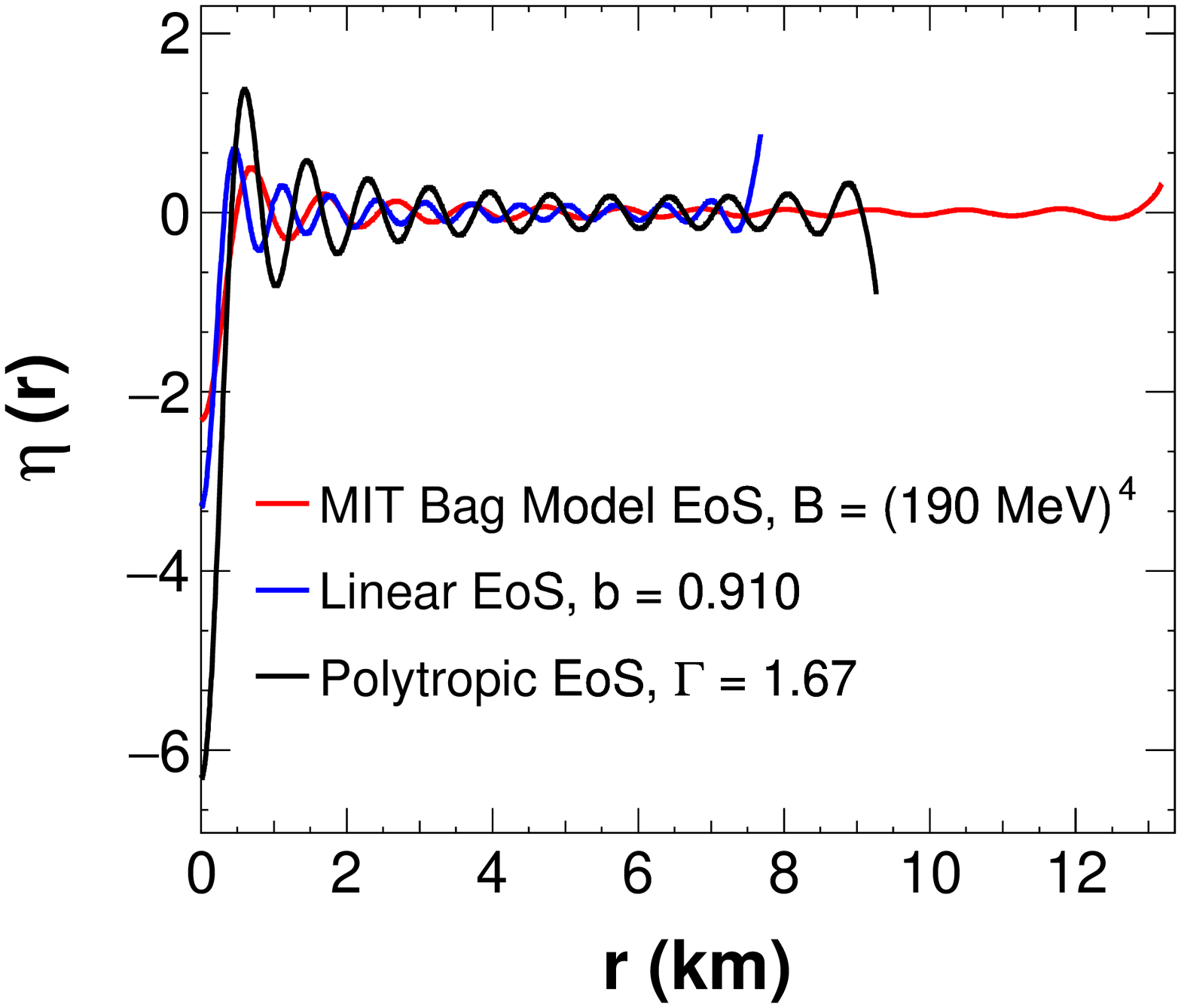}}
        \vspace{-0.3cm}
        \caption{Variation of pressure perturbation parameter $\eta(r)$ with
        radial distance of SS for three EoSs. The left plot is for the $f$-mode
        and the other plot is for the $p_{22}$-mode.}
    \label{fig11}
    \end{figure*}

\begin{figure*}
        \centerline{
        \includegraphics[scale = 0.3]{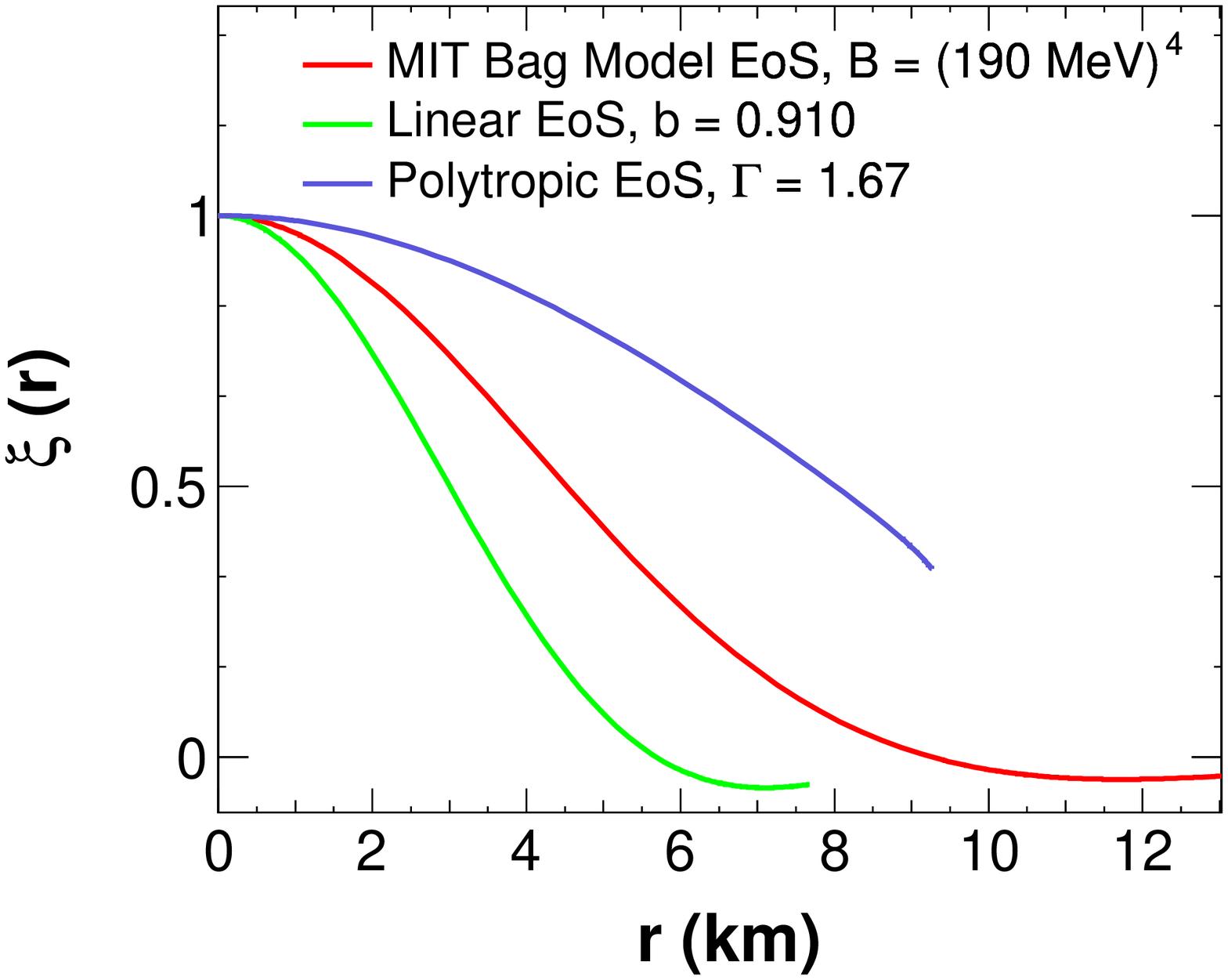}\hspace{0.5cm}
        \includegraphics[scale = 0.3]{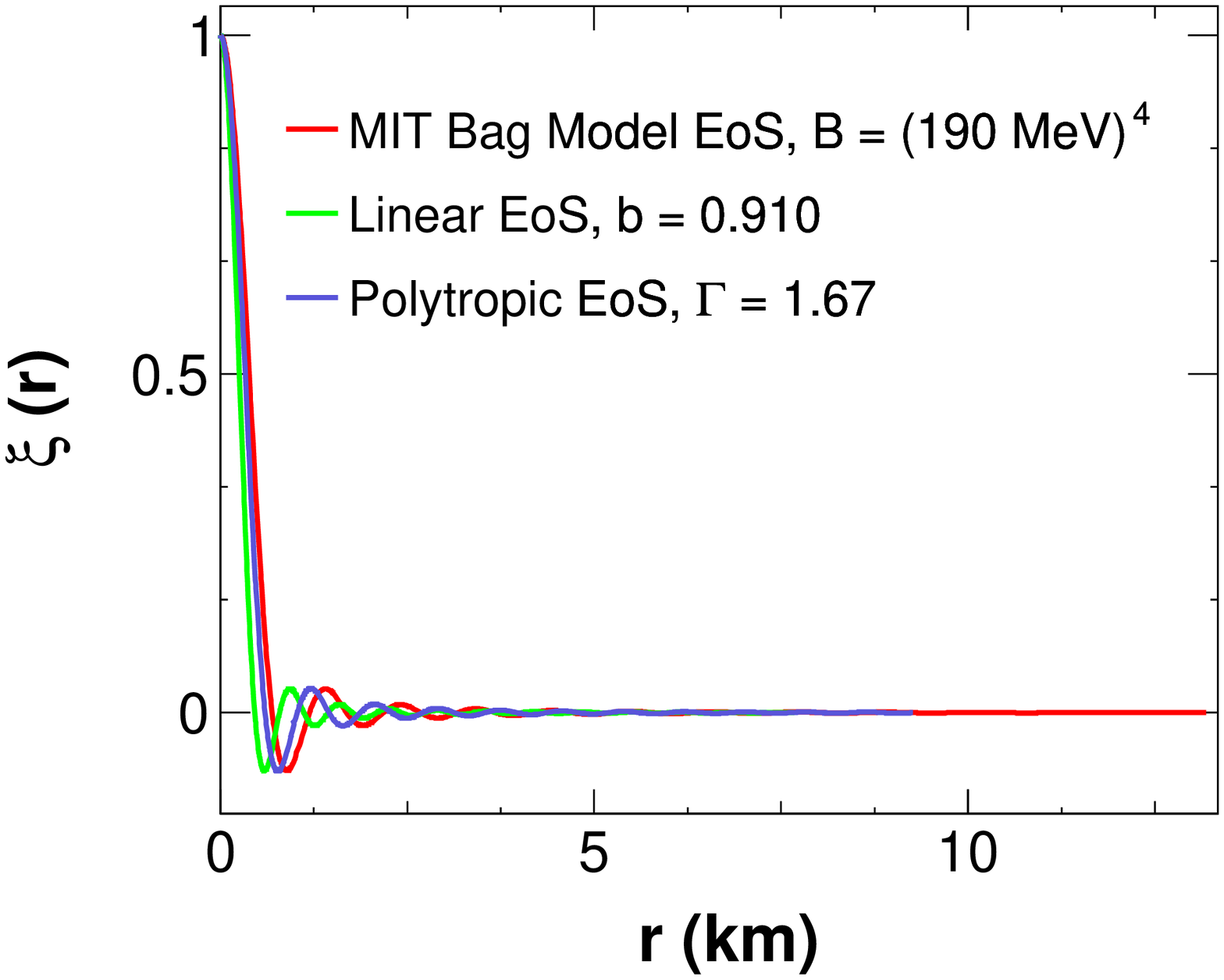}}
        \vspace{-0.3cm}
        \caption{Variation of radial perturbation parameter $\xi(r)$ with
        radial distance of SS for three EoSs. The left plot is for the $f$-mode
        and the right plot is for the $p_{22}$-mode.}
    \label{fig12}
    \end{figure*}

As a comparative analysis of the predictions of all three models considered
in this work, we have shown the variation of pressure and radial perturbations
for these three model EoSs in Fig.\ \ref{fig11} and Fig.\ref{fig12}
respectively. In the left plot of Fig.\ \ref{fig11} the comparison of variation
of $f$-mode of pressure perturbations with the star radial distance is
represented. The right plot of this figure is shown the same for the  $p_{22}$
oscillation mode of the star. As clear from this figure, the linear EoS is
predicting a star of smallest radius. For all these three EoSs, the
pattern of variation of $\eta_n(r)$ is same, maximum near the centre and near
the surface of the star.
In Fig.\ \ref{fig12}, the radial perturbations are compared for these three
different pressure-energy density relations. In the left plot the results of
$f$-mode and in the right plot the results of $p_{22}$-mode are
compared. Unlike the pressure perturbations, the radial perturbation values are
found to maximum near the centre of each stars and are found to follow the same
pattern in all EoSs. In this case also the linear EoS gives the star with the
smallest radius.

\begin{figure*}
	\centerline{
	\includegraphics[scale=0.35]{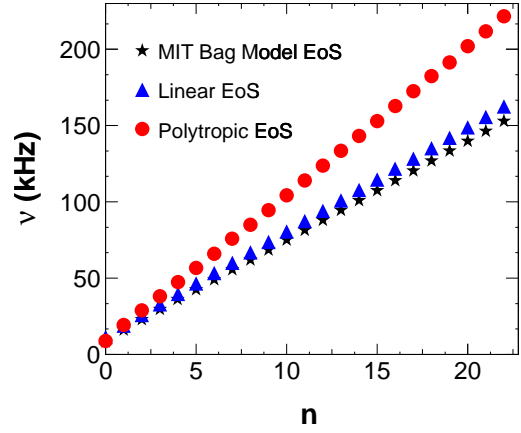}}
	\vspace{-0.3cm}
	\caption{Comparative variation of radial oscillation frequencies 
        $\nu_{n}$ with respect to oscillation modes (n) for the MIT Bag model 
        EoS (black point) with $B=(243\,\mbox{MeV})^{4}$, linear EoS 
        (blue points) with $b=0.910$ and polytropic EoS (red points) with 
        $\Gamma=2$. These are maximum frequencies for each mode
        given by these EoSs for these respective constant parameters.} 
	\label{fig2} 
	\end{figure*}

Table \ref{tab:table2} shows all calculated radial oscillation frequencies of 
such stars 
for the aforementioned EoSs obtained by using all three chosen constant 
parameters in each EoS. Here the oscillation frequencies of respective 22 
modes from the $f$-mode (in order $n$) are given in kHz. It is seen that 
for the MIT bag model EoS and the polytropic EoS, the oscillation frequencies 
of all modes increase with the increasing values of constant parameters, 
while the situation is opposite in the case of linear EoS. Thus, individually 
frequencies are maximum for $B=(243\, \mbox{MeV})^{4}$ of MIT Bag model EoS,
for $b=0.910$ of linear EoS and for $\Gamma=2$ of polytropic EoS. But, among 
these
EoSs polytropic EoS with $\Gamma=2$ gives highest maximum frequencies of radial
oscillations. Moreover, for the purpose of visualization we have shown in Fig.\ 
\ref{fig2} the graphical representation of comparison of maximum radial 
oscillation frequencies obtained for each EoSs with the said respective 
constant parameters.   
\begin{table*}
	\caption{\label{tab:table2} Radial oscillation frequencies $\nu_{n}$ 
        in kHz for three EoSs. Rows under each EoS are for smaller to bigger
        values of respective constant parameter. These are 
        $B = (190\,\mbox{MeV})^4$, $(217\,\mbox{MeV})^4$ and 
        $(243\,\mbox{MeV})^4$ for the MIT Bag model EoS; $b = 0.910$, $0.918$ 
        and $0.926$ for the linear EoS; and $\Gamma=1.5$, $1.67$ and $2$ for 
        the polytropic EoS.\vspace{5pt}}
\begin{tabular}{c|ccc|ccc|ccc}
\hline \hline
Modes (Order n) &&\hspace{-12pt} MIT Bag model EoS \hspace{-12pt} &&& \hspace{3pt} Linear EoS \hspace{3pt} &&& Polytropic EoS\\[5pt]
\hline
$f$      (0)  & 5.08  & 6.58   & 8.27    & 11.44  & 10.86  & 10.28  & 1.57   &  4.71  & 8.73 \\
$p_{1}$  (1)  & 9.88  & 12.79  & 16.08   & 18.85  & 17.98  & 17.12  & 9.26   & 12.48  & 19.28 \\
$p_{2}$  (2)  & 14.08 & 18.24  & 22.93   & 25.88  & 24.72  & 23.57  & 14.56  & 18.84  & 28.85 \\
$p_{3}$  (3)  & 18.15 & 23.50  & 29.54   & 32.78  & 31.34  & 29.89  & 19.57  & 24.90  & 38.15 \\
$p_{4}$  (4)  & 22.16 & 28.70  & 36.08   & 39.63  & 37.90  & 36.16  & 24.46  & 30.99  & 47.37\\
$p_{5}$  (5)  & 26.15 & 33.87  & 42.58   & 46.45  & 44.43  & 42.41  & 29.27  & 37.25  & 56.63 \\
$p_{6}$  (6)  & 30.14 & 39.04  & 49.07   & 53.26  & 50.96  & 48.65  & 34.05  & 43.64  & 65.99 \\
$p_{7}$  (7)  & 34.12 & 44.19  & 55.55   & 60.07  & 57.47  & 54.88  & 38.81  & 50.13  & 75.44 \\
$p_{8}$  (8)  & 38.10 & 49.35  & 62.03   & 66.88  & 63.99  & 61.11  & 43.57  & 56.68  & 84.98\\
$p_{9}$  (9)  & 42.09 & 54.51  & 68.52   & 73.90  & 70.51  & 67.35  & 48.36  & 63.26  & 94.59 \\
$p_{10}$ (10) & 46.07 & 59.67  & 75.00   & 80.50  & 77.04  & 73.58  & 53.18  & 69.87  & 104.26\\
$p_{11}$ (11) & 50.05 & 64.83  & 81.49   & 87.31  & 83.57  & 79.82  & 58.02  & 76.48  & 113.96\\
$p_{12}$ (12) & 54.04 & 69.99  & 87.97   & 94.13  & 90.10  & 86.07  & 62.89  & 83.11  & 123.69\\
$p_{13}$ (13) & 58.02 & 75.15  & 94.46   & 100.95 & 96.63  & 92.32 & 67.77   & 89.75  & 133.44\\
$p_{14}$ (14) & 62.01 & 80.32  & 100.95  & 107.78 & 103.17 & 98.57 & 72.68   & 96.39  & 143.20\\
$p_{15}$ (15) & 66.00 & 85.48  & 107.44  & 114.61 & 109.71 & 104.82 & 77.60  & 103.03 & 152.98 \\
$p_{16}$ (16) & 69.98 & 90.64  & 113.93  & 121.44 & 116.26 & 111.08 & 82.52  & 109.68 & 162.77\\
$p_{17}$ (17) & 73.97 & 95.81  & 120.42  & 128.28 & 122.81 & 117.34 & 87.46  & 116.32 & 172.57\\
$p_{18}$ (18) & 77.96 & 100.97 & 126.91  & 135.12 & 129.36 & 123.61 & 92.40  & 122.91 & 182.38\\
$p_{19}$ (19) & 81.95 & 106.14 & 133.41  & 141.96 & 135.91 & 129.87 & 97.35  & 129.62 & 191.46\\
$p_{20}$ (20) & 85.94 & 111.30 & 139.90  & 148.80 & 142.27 & 136.14 & 102.30 & 136.29 & 202.00\\
$p_{21}$ (21) & 89.92 & 116.47 & 146.39  & 155.65 & 149.03 & 142.41 & 106.80 & 142.92 & 211.82\\
$p_{22}$ (22) & 93.91 & 121.64 & 152.89  & 162.50 & 155.59 & 148.68 & 111.66 & 149.60 & 221.64\\[1pt]
\hline \hline
\end{tabular}
\end{table*}
\hspace{-12pt}
This table as well as figure show that, with increasing number of modes, the 
values of radial oscillation frequencies are getting larger. From the table 
it is also 
clear that in case of the maximum radial oscillation frequencies, the frequency 
differences between consecutive modes are larger for the polytropic EoS than 
that for the linear and MIT Bag model EoSs. Otherwise, the linear EoS gives
wider frequency differences. In all cases, the MIT Bag model EoS gives the 
smallest difference between frequencies. The details about all these behaviours
of radial oscillation frequencies in each EoS are discussed as follows. 

\begin{figure*}
	\centerline{
	\includegraphics[scale=0.27]{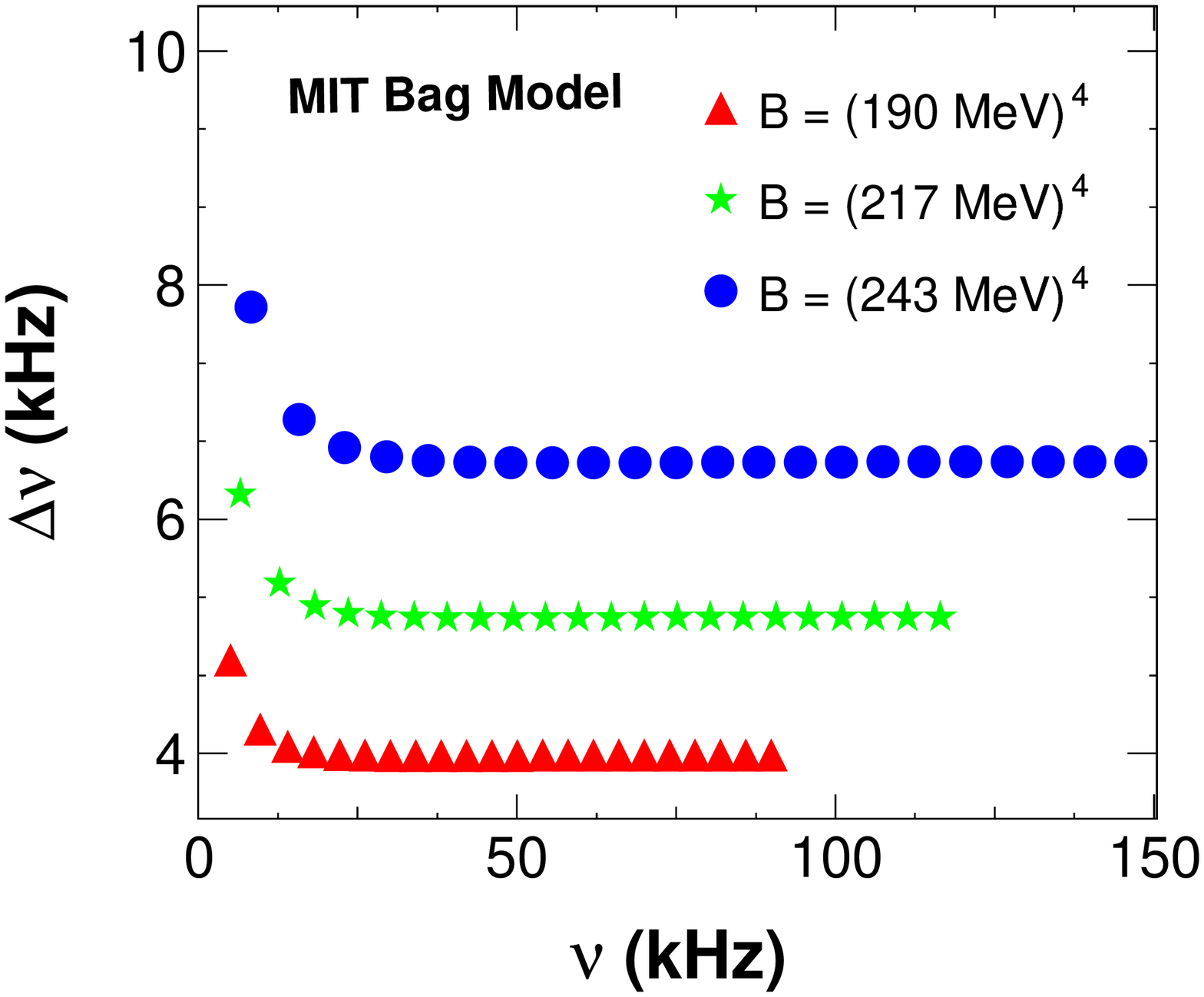}
        \hspace{0.2cm}
	\includegraphics[scale=0.27]{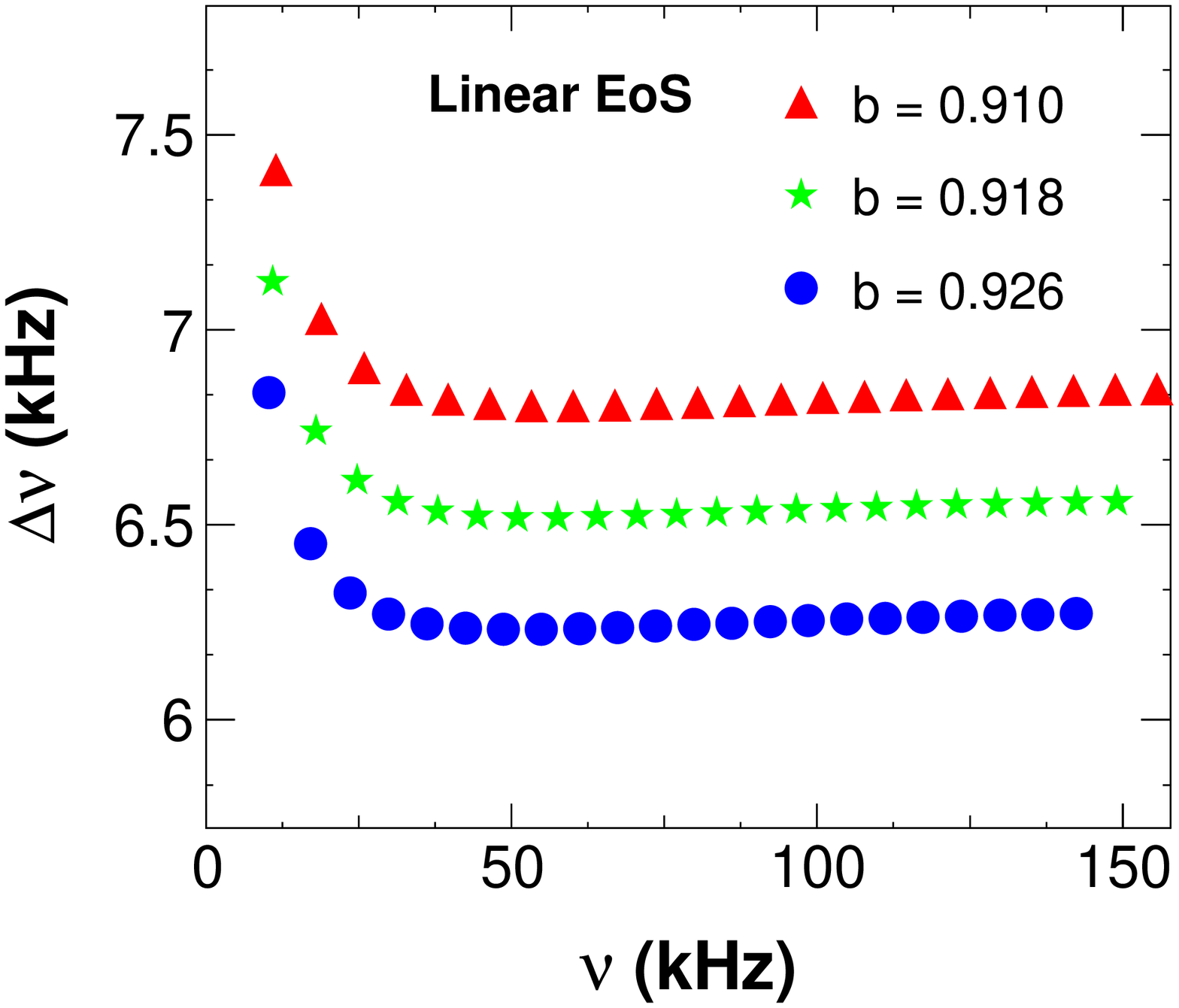}
        \hspace{0.2cm}
        \includegraphics[scale=0.27]{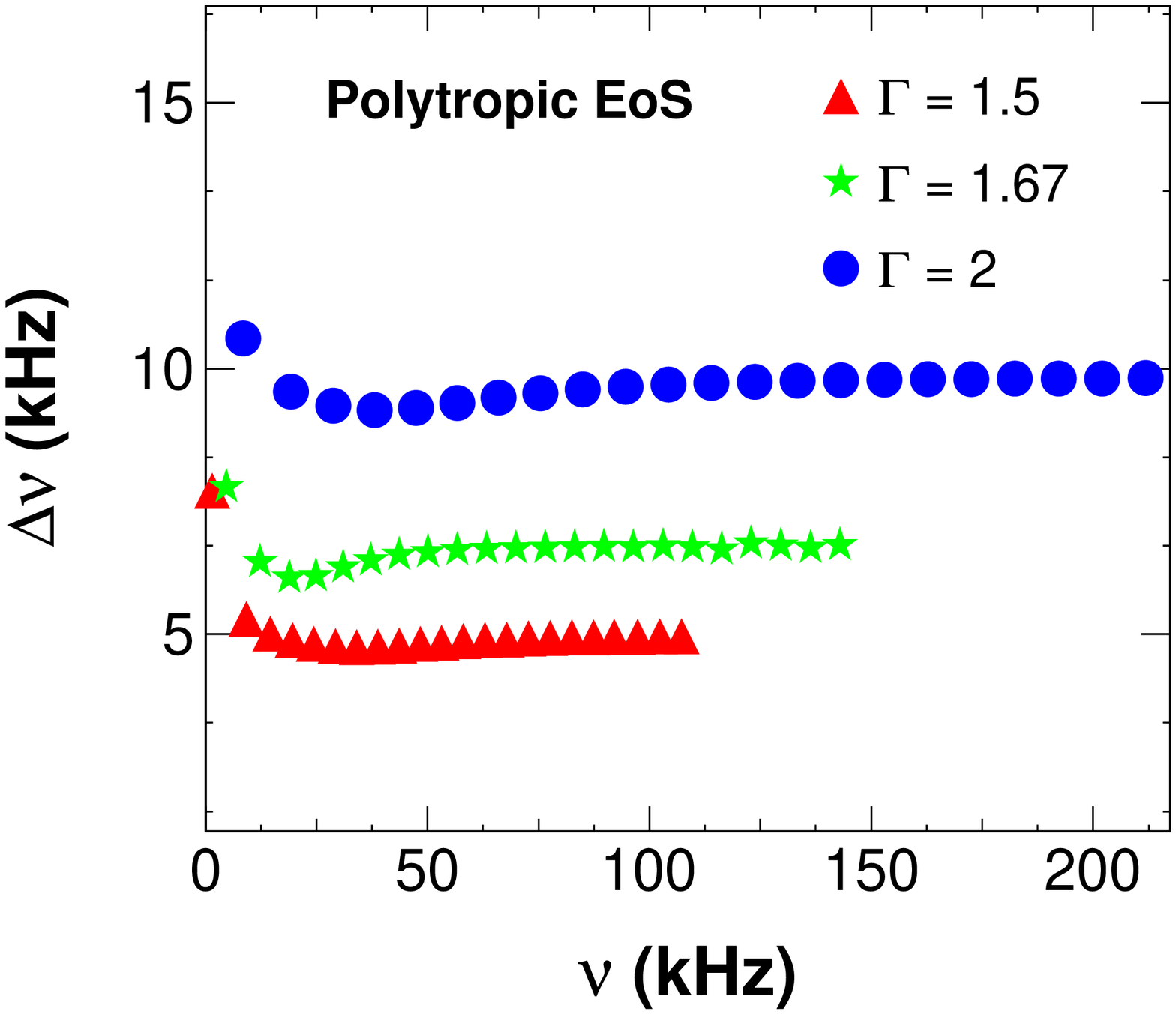}
        }
        \vspace{-0.3cm}
	\caption{Variation of radial oscillation frequency differences between 
        consecutive modes with the mode frequency $\nu_{n}$ for the (a) MIT 
        Bag Model EoS [left plot], (b) linear EoS [middle plot] and (c) 
        polytropic EoS [right plot].} 
    \label{fig3}
	\end{figure*}

\begin{figure*}
	\centerline{
	\includegraphics[scale=0.35]{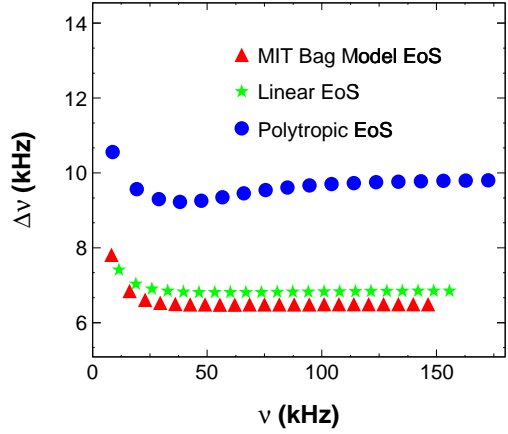}}
    \vspace{-0.3cm}
	\caption{Comparison of frequency differences between consecutive modes 
        with respect to the mode frequency for all three EoSs with the
        model parameters for which respective EoSs are giving maximum 
       oscillation frequencies.} 
    \label{fig4}
	\end{figure*}

The dependence of radial oscillation frequencies on the model parameters in 
terms of frequency difference between consecutive modes is shown in Fig.\ 
\ref{fig3} and \ref{fig4}. These figures are for frequency differences with 
respect to oscillation frequencies in considered EoSs. For MIT Bag model EoS, 
with three values of Bag constants, viz., $B=(190\, \mbox{MeV})^4$, $(217\,
\mbox{MeV})^4$ and $(243\, \mbox{MeV})^4$, the dependence of frequencies is 
shown in the left plot of Fig.\ \ref{fig3}. The highest limiting value of Bag 
constant $B=(243\,\mbox{MeV})^4$ is giving the larger values of oscillation 
frequency differences. Whereas for $B=(190\, \mbox{MeV})^4$, the 
frequency differences are found to be much smaller than that for the two other 
values. This indicates that when the values of Bag constant $B$ increases, the 
radial oscillation frequency difference between two consecutive modes rises
significantly. Again, for each value of constant $B$, the separation between 
two consecutive modes decreases towards higher order modes. Thus this 
separation is maximum for the $f$-mode, followed by lower order $p$-modes. For 
values of $B$ lying in between $(190\,\mbox{MeV})^4$ and $(243\,\mbox{MeV})^4$, 
the intermediate values of the oscillation frequencies are acquired. Similar 
behaviours of oscillation frequencies are obtained for linear EoS with three 
values of linear constants: $b=0.910$, $b=0.918$ and $b=0.926$, which is shown 
in the middle plot of Fig.\ \ref{fig3}. However, in this case for the larger 
value of linear constant, i.e.\ for $b=0.926$, the frequency differences are 
found to be smaller than that for the other two smaller values of $b$. For the 
lowest value of $b$, i.e.\ for $b=0.910$, we get highest values of frequency
differences and $b=0.918$ gives the intermediate values. For each values of 
$b$ considered here, the separation 
between the consecutive modes is maximum for $f$-mode and is gradually getting 
smaller for higher order modes. Same study is made for the polytropic EoS with 
polytropic exponent $1.5$, $1.67$ and $2$, which are shown in the right plot of 
Fig.\ \ref{fig3}. Among these considered values of polytropic exponent, we have 
obtained maximum oscillation frequency differences for $\Gamma=2$. Similar to 
the other EoSs, for this model also the frequency differences are getting 
smaller for higher order modes. However for $\Gamma=1.67$ a slight variation 
in the pattern is observed than that for the other two considered values. Like 
MIT Bag model EoS, for this EoS also, we have obtained that, larger is the 
value of $\Gamma$ higher is the oscillation frequency differences. 
In Fig.\ \ref{fig4}, a comparison between these EoSs are 
made. The frequency differences for linear and MIT Bag model EoSs are 
decreasing gradually with increasing frequencies. For polytropic EoS, the 
pattern is slightly distorted as shown in figure. Maximum values of frequency
differences are observed for polytropic EoS with $\Gamma=2$, whereas
minimum differences are observed for MIT Bag model EoS with 
$B = (243\,\mbox{MeV})^4$ but these are comparable to that of linear EoS with 
$b = 0.910$.  Thus it is clear that, the radial oscillation frequencies of SSs 
crucially depend on the model and the model parameters used. 
\subsection{GWE frequencies of SSs}
\begin{table*}
\caption{\label{tab:table3} GWE frequencies in kHz from SSs for three EoSs with
different model parameters.}

\begin{ruledtabular}
\begin{tabular}{ccccc}
EoSs & Model  & Echo & GWE & GWE \\[-3pt] & Parameter & time (ms) &  Frequency 
(kHz) & Repetition Frequency (kHz) \\ \hline
\multirow{3}{*}{MIT Bag model} 
& $B=(190\, \mbox{MeV})^{4}$ & 0.078 & 39.91 & 6.35 
\\ & $B=(217\, \mbox{MeV})^{4}$ & 0.060 & 51.70 & 8.23
\\ & $B=(243\, \mbox{MeV})^{4}$ & 0.048 & 64.98 & 10.34\\
\multirow{3}{*}{Linear EoS} 
& $b=0.910 $ & 0.043 & 72.90 & 11.60
\\ & $b=0.918$ & 0.044 & 70.21 & 11.18 
\\ & $b=0.926$ & 0.046 & 67.42 & 10.73 \\ 
\end{tabular}
\end{ruledtabular}
\end{table*}
The GWE frequencies of different SS models are given in Table \ref{tab:table3}. 
The possible SS models which fulfil the criterion for emitting GWE 
frequencies, i.e.\ featuring a photon sphere and the radius not exceeding the 
Buchdahl's limit radius are shown in Fig.\ \ref{fig1}. So, the EoSs are 
chosen in order to obtain maximum mass ranges of the stellar configuration. 
Along with the two models as shown in the Fig.\ \ref{fig1}, which have the 
required compactness, we have chosen four other EoSs by varying the model 
parameter of MIT Bag model and linear EoS. For MIT Bag model [Eq.\ \eqref{eq1}] 
with $B={(190\,\mbox{MeV})}^4, {(217\,\mbox{MeV})}^4$ and ${(243\,\mbox{MeV})}
^4$, and linear EoS [Eq.\ \eqref{eq2}] with $b=0.910, 0.918$ and $0.926$, the 
compactness within the range of $0.33$ to $0.44$ can be obtained. So we have 
calculated GWE frequencies in these models.
\begin{figure*}
        \centerline{
        \includegraphics[scale = 0.3]{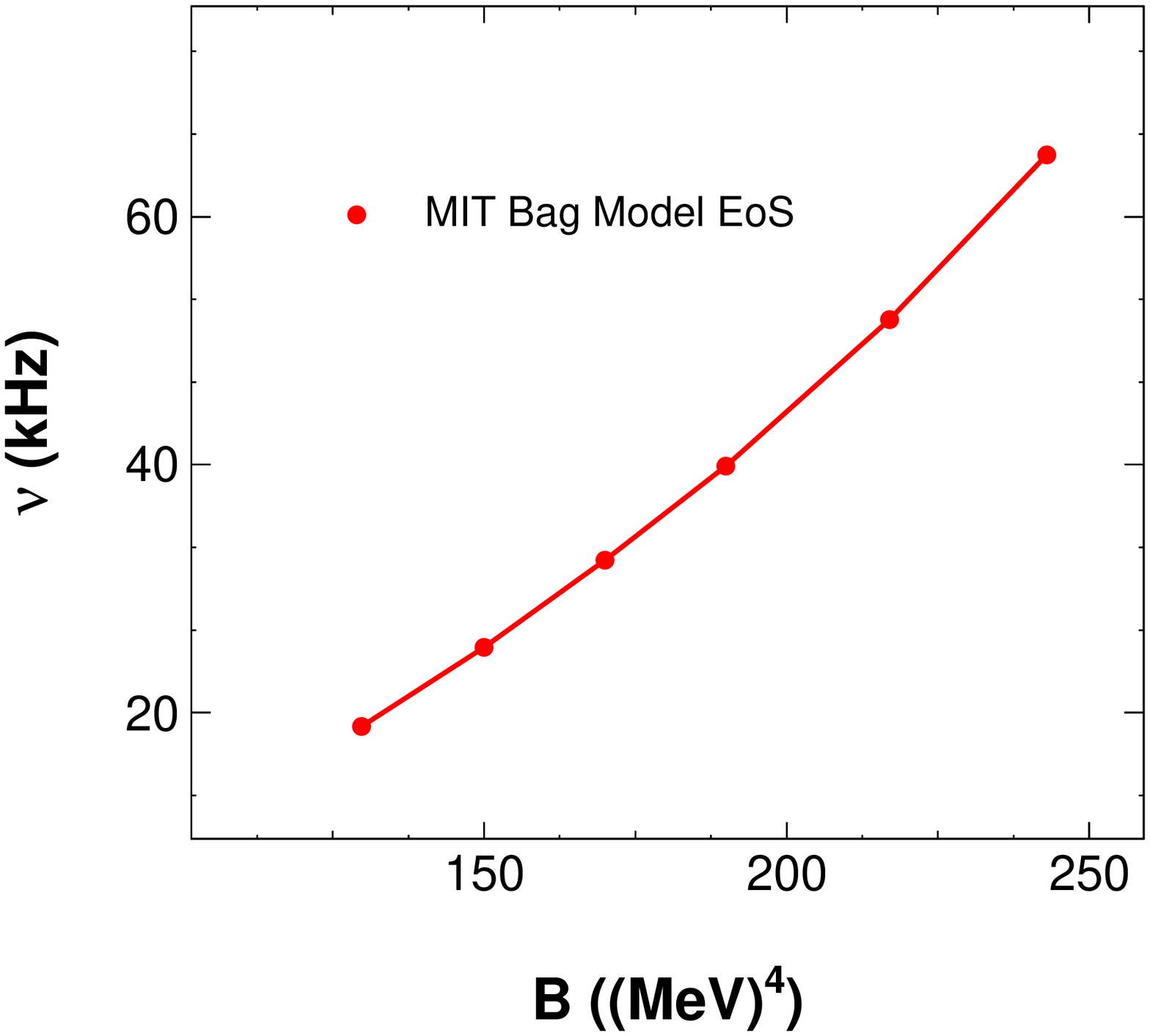}\hspace{0.5cm}
        \includegraphics[scale = 0.3]{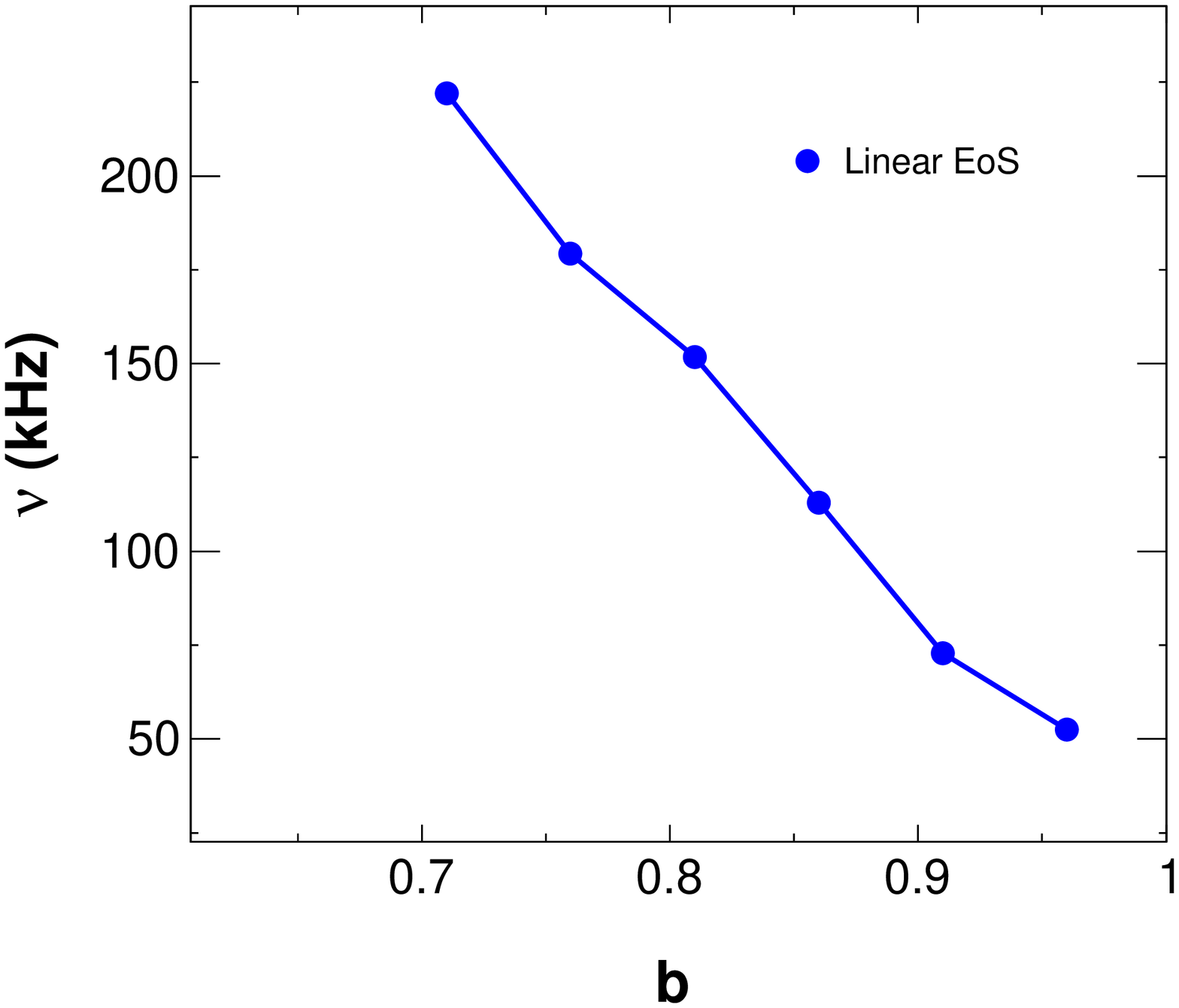}}
        \vspace{-0.0cm}
        \caption{Variation of GWE frequencies with different model 
parameters: left plot is for MIT Bag model with different Bag constant $B$ and 
right plot is for linear EoS with linear constant $b$.}
    \label{fig13}
    \end{figure*} 

Here we report the GWE frequencies obtained for three values of Bag constant 
$B$: $(190\,\mbox{MeV})^{4}$, $(217\, \mbox{MeV})^{4}$ and $(243\, 
\mbox{MeV})^{4}$ and three values of linear constant $b$: $0.910$, $0.918$ and 
$0.926$. These considered EoSs lead to different types of SSs with varying 
maximum masses. For MIT Bag model, the maximum masses are $M_{max} 
\approx 3.29\, M_{\odot}$ with $B=(190\, \mbox{MeV})^{4}$, $M_{max} 
\approx 2.54\, M_{\odot}$ with $B=(217\, \mbox{MeV})^{4}$ and $M_{max}
\approx 2.02\, M_{\odot}$ with $B=(243\, \mbox{MeV})^{4}$. The linear EoS leads 
to maximum masses $M_{max}\approx 1.77\, M_{\odot}$ for $b=0.910$, $M_{max}
\approx 1.84\, M_{\odot}$ for $b=0.918$ and $M_{max}\approx 1.92\, M_{\odot}$ 
for $b=0.926$. Thus, it is seen that with the increasing value of 
$B$ of MIT Bag model EoS, the maximum mass of SS decreases, whereas it 
increases with the increasing value of $b$ of linear EoS.
Our result shows that the GWE frequency increases with increase in values of 
Bag constant $B$. For $B={(190\,\mbox{MeV})}^4$, the GWE frequency is $
\omega_{MIT,\, echo}= 39.91$ kHz which is much smaller than the values 
corresponding to $B={(217\,\mbox{MeV})}^4$ and ${(243\,\mbox{MeV})}^4$, which 
also show a large difference ($\sim 13.28$ kHz). Thus GWE frequencies for these 
SS models show a large variation in their values. For linear EoSs, with 
$b=0.910$, maximum echo frequency of $72.90$ kHz is observed. For $b=0.918$, 
the echo frequency is $\approx 67.42$ kHz, which is near to the value of 
$\omega_{MIT,\, echo}$ with $B={(243\,\mbox{MeV})}^4$. The echo frequency 
for polytropic EoS is not observed as the required compact SSs are not found 
for such type of EoSs. Along with the characteristic echo frequencies the 
repetition frequencies of GWEs are also calculated (shown in 
Table \ref{tab:table3} ). The repetition frequencies are found to be 
much smaller than the GWE frequencies and they are following the same 
pattern as the corresponding echo frequencies. 
    
Further, for more clarity the variation of GWE frequencies with different 
model patrameters for MIT Bag model and linear EoSs are shown in Fig.\ \ref{fig13}. The first plot of this figure shows that, with increasing value of 
Bag constant $B$, GWE frequency increases. Whereas from the second plot it is 
clear that unlike MIT Bag model, GWE frequency decreases with the increasing 
value of linear constant $b$.


\section{Summary and Conclusions}\label{conclusion} 
In the first part of this study, we have computed frequencies of 22 lowest 
radial oscillation modes and in the second part, we have computed the echo 
frequencies of SSs considering that the fluid pressure follows certain EoSs. 
Considering isotropic configuration of strange matter, we have integrated the 
TOV equations in general relativistic case numerically to get the values of 
mass, radius of the star and the metric term. For the defined pressure-energy 
density relations, we have obtained different SS configurations. After that, 
the radial and pressure perturbation equations respectively in $\xi=\Delta r/r$ 
and $\eta=\Delta p/p$ are solved for the eigenfrequencies. To calculate echo 
frequencies those stars are chosen which can echo GWs by requiring the 
condition for compactness. The typical echo time is then calculated and the 
characteristic GWE frequencies along with the repetition frequencies are 
determined for linear and MIT Bag model EoSs. 

In this study of radial oscillation modes of SSs,  we have calculated the 
respective $f$-mode and $p$-modes of oscillation frequencies associated with 
different SSs. The characteristic echo frequencies and the respective 
repetition of the echo frequencies are also calculated for SSs. Table 
\ref{tab:table2} and Table \ref{tab:table3} summarise the results of the 
work. From the numerical results we can summarise our work as follows. 

The amplitude of radial perturbations $\xi_{n}(r)$ is larger 
closer to the centre and much smaller near the surface for all the three 
EoSs. For all mode of frequencies the maximum amplitude of $\xi_{n}(r)$ is 
same. The pressure perturbation $\eta_{n}(r)$ is larger closer to the 
centre and near the surface of the star for MIT Bag model, linear EoS and 
polytropic EoS. The values of $ \xi_{n}(r)$ and $\eta_{n}(r)$ are differing 
from model to model and also for parameters of the models. 

Polytropic EoS gives the largest value of radial frequencies among 
the three EoSs. MIT Bag model and linear EoSs have nearer values of radial 
frequencies. With increasing modes, the frequencies also increases. The 
separation between two consecutive oscillation modes decreases with the 
increase in frequency or mode for all three EoSs. However, this effect is less 
pronounced for higher order modes. Moreover, the frequency differences for 
polytropic EoS have shown a slight distortion. Further, the magnitude of
this frequency separation depends upon the EoS as well as on the associated 
model parameter. Among all three EoSs the polytropic EoS with $\Gamma = 2$
gives the maximum separation. Thus from these results, it is clear that 
oscillation frequencies show high dependency on model and model parameters. 

In case of GWE frequency not all EoSs are able to emit echo. Those EoSs which 
give stellar configuration with much higher compactness are able to give echo 
frequencies. The echo frequency obtained for MIT bag model and linear EoSs 
show a distinct variation in their values depending on the model parameters. 
Thus it can be said that, the calculated characteristic echo 
frequencies as well as repetition frequencies changes with model and model 
parameters.

The existence of SS has not based on firm footing till now, despite of the 
possible evidences of SS candidates. As mentioned in above sections, the 
structure of SS crucially depends on the EoS. Considering some relevant EoSs 
for such stars one will get different SS configurations. The knowledge of 
possible SS configurations will help in searching such compact stars. Again 
from the reflected GWE signal it is also possible to say about the structural 
behavior of such stars. The prediction of echo signal will get its firm 
foundation by the experimental detection of such signal using some future 
generation of GW detectors. We think with a sufficient amount of 
experimental data on echo frequencies of GWs, it could be possible to 
constrain the parameters of EoSs and hence to find out the most viable EoS. 
In this study the echo frequencies of GWs are  found to be above 30 kHz range. 
Advanced LIGO, Advanced Virgo and KAGRA are projected to have sensitive to 
GWs with frequencies of $\sim$ 20 Hz - 4 kHz and with amplitudes of 
$\sim 2\times 10^{-22}$ - $4 \times 10^{-24}$ strain/$\sqrt{\mbox{Hz}}$ 
\citep{martynov, abbot}. Due to limited by shot noise at high 
frequencies, currently LIGO and Virgo observatories have a sensitivity of 
$\ge 2\times 10^{-23}$ strain/$\sqrt{\mbox{Hz}}$ at $3$ kHz. According to D. 
Martynov et al. this sensitivity can be enhanced by an optical configuration 
of detectors using the current interferometer topology to reach 
$\ge 7\times 10^{-25}$ strain/$\sqrt{\mbox{Hz}}$ at $2.5$ kHz. These proposed 
instruments with optimal arm length of $\approx 20$ km would have the 
sensitivity to detect the amount of postmerger neutron star oscillations at 
per the third generation detectors, such as Cosmic Explorer (CE)
\citep{abbott} and Einstein Telescope (ET) \citep{punturo}. The 
CE is a proposed 40 km arm length L-shaped observatory to deepen the GW view 
of the cosmos, whose sensitivity may reach below $10^{-25}$ 
strain/$\sqrt{\mbox{Hz}}$ at above few kHz frequencies. On the other hand, ET 
is a $10$ km arm length L-shaped underground proposed observatory which will 
be able to reach the sensitivity of $> 3\times10^{-25}$ 
strain/$\sqrt{\mbox{Hz}}$ at $100$ Hz and of $\sim 6\times10^{-24}$ 
strain/$\sqrt{\mbox{Hz}}$ at $\sim 10$ kHz. It is clear that none of present 
or near future generation of GW detectors could reach a sensitivity required 
to detect GWEs in the range of our study. However, in a very recent study by
S. L. Danilishin et al. \citep{danilishin} shows that by the application 
of advanced quantum techniques to suppress the quantum noise at high frequency 
end in the design of GW detectors, the sensitivity of the present GW detectors 
can be enhanced significantly. So, the
application of such techniques to the proposed third generation of detectors
mentioned above may lead to increase their sensitivity to our expected level.
Otherwise, we have to wait for the distant future, fourth generation GW
detectors with sufficiently higher sensitivity in the kHz range of frequencies
to test the results of this study. Obviously, a detailed study on the
possibility of detection GWE frequencies and ways to increase detector
sensitivities can shed more light in this regard in future.

\section*{Acknowledgments}
JB is grateful to Dibrugarh University, for the financial support 
through the grant `DURF-2019-20' while carrying out this work. She also shows 
her sincere gratitude towards D. J. Gogoi for useful discussion. UDG is 
thankful to the Inter-University Centre for Astronomy and Astrophysics
(IUCAA), Pune for hospitality during his visits to the institute under the 
Visiting Associateship program.


\bibliographystyle{apsrev}
\end{document}